\documentclass[sigconf,screen]{acmart}

\usepackage{xr}
\usepackage{multirow}
\usepackage{caption}
\usepackage{subcaption}
\usepackage{graphicx}
\usepackage{cleveref}
\usepackage{xcolor}
\usepackage{soul}
\usepackage{color, colortbl}
\definecolor{Gray}{gray}{0.7}

\newlength\myboxwidth
\AtBeginDocument{%
  \providecommand\BibTeX{{%
    \normalfont B\kern-0.5em{\scshape i\kern-0.25em b}\kern-0.8em\TeX}}}

\copyrightyear{2024}
\acmYear{2024}
\setcopyright{rightsretained}
\acmConference[CHI '24]{Proceedings of the CHI Conference on Human Factors in Computing Systems}{May 11--16, 2024}{Honolulu, HI, USA}
\acmBooktitle{Proceedings of the CHI Conference on Human Factors in Computing Systems (CHI '24), May 11--16, 2024, Honolulu, HI, USA}
\acmDOI{10.1145/3613904.3642522}
\acmISBN{979-8-4007-0330-0/24/05}

\begin{document}

\title{Designing Distinguishable Mid-Air Ultrasound Tactons with Temporal Parameters}



\author{Chungman Lim}
\email{chungman.lim@gm.gist.ac.kr}
\orcid{0000-0002-7857-3322}
\affiliation{%
  \institution{Gwangju Institute of\\ Science and Technology}
  \country{Republic of Korea}}

\author{Gunhyuk Park}
\email{maharaga@gist.ac.kr}
\orcid{0000-0003-2677-5907}
\affiliation{%
  \institution{Gwangju Institute of\\ Science and Technology}
  \country{Republic of Korea}
}

\author{Hasti Seifi}
\email{hasti.seifi@asu.edu}
\orcid{0000-0001-6437-0463}
\affiliation{%
  \institution{Arizona State University}
  \city{Tempe}
  \country{United States}}



\begin{abstract}

Mid-air ultrasound technology offers new design opportunities for contactless tactile patterns (i.e., Tactons) in user applications. Yet, few guidelines exist for making ultrasound Tactons easy to distinguish for users. In this paper, we investigated the distinguishability of temporal parameters of ultrasound Tactons in five studies (n=72 participants). Study 1 established the discrimination thresholds for amplitude-modulated (AM) frequencies. In Studies 2--5, we investigated distinguishable ultrasound Tactons by creating four Tacton sets based on mechanical vibrations in the literature and collected similarity ratings for the ultrasound Tactons. We identified a subset of temporal parameters, such as rhythm and low envelope frequency, that could create distinguishable ultrasound Tactons. Also, a strong correlation (mean Spearman's $\rho$=0.75) existed between similarity ratings for ultrasound Tactons and similarities of mechanical Tactons from the literature, suggesting vibrotactile designers can transfer their knowledge to ultrasound design. We present design guidelines and future directions for creating distinguishable mid-air ultrasound Tactons.

\end{abstract}




\begin{teaserfigure}
  \centering
  \includegraphics[width=\textwidth]{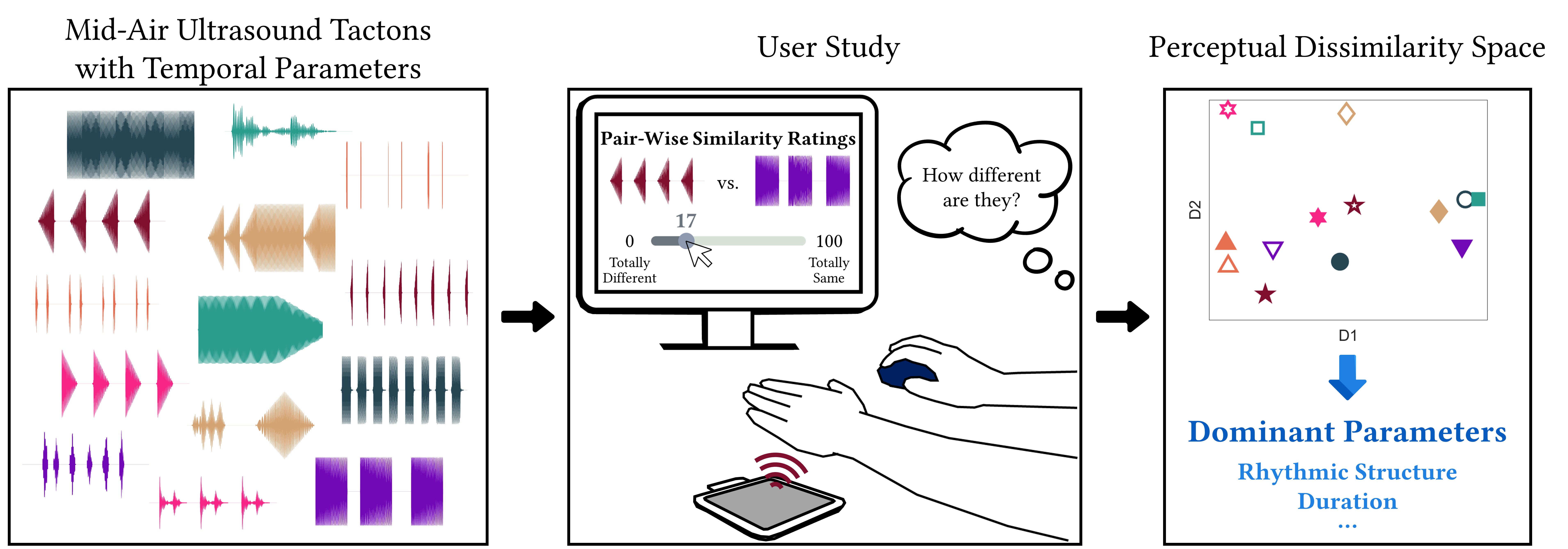}
  \caption{ 
   Overview of our procedure for identifying the distinguishable temporal parameters of mid-air ultrasound Tactons.
   We created mid-air ultrasound Tactons using temporal parameters. 
   People felt all pairs of the Tactons using the ultrasound haptic device and rated the perceptual similarity of the Tactons.
   Finally, we derived perceptual dissimilarity spaces and identified the dominant parameters for distinguishable ultrasound Tactons.
  }
  \label{fig:teaser}
  \Description{This figure illustrates the procedure of the similarity rating study. A mid-air ultrasound Tacton (i.e., tactile pattern) set is designed by varying temporal parameters. Users use an ultrasound device and a GUI program to assess pair-wise perceptual similarity for Tactons in this set. Based on the perceptual dissimilarity space derived from user ratings, dominant parameters are identified that contribute to the distinguishability of ultrasound Tactons.}
\end{teaserfigure}

\maketitle

\section{Introduction}

Ultrasonic mid-air haptic technology presents new possibilities for delivering contactless haptic feedback in various user applications.
The technology can vibrate human skin from a distance by focusing ultrasound waves in mid-air~\cite{rakkolainen2020survey}. To create different sensations, designers can modulate the parameters of the ultrasound pattern over time~\cite{obrist2013talking,wojna2023exploration} or move the ultrasound focal point in space to render a shape on the user's skin~\cite{long2014rendering,hajas2020mid}. The resulting mid-air ultrasound patterns, i.e., tactile icons or Tactons, can communicate information or emotion to users in various applications such as touchless interactions with public displays~\cite{vi2017not, limerick2019user}, automotive user interfaces~\cite{harrington2018exploring}, and virtual reality environments~\cite{hwang2017airpiano, howard2022ultrasound}.

To convey information, Tactons must be easy to distinguish for users.
A designer may create Tactons by systematically varying the parameters of a haptic signal (parameter-based approach)~\cite{yoo2015emotional, ternes2008designing}.
Alternatively, the designer may create a Tacton to remind users of metaphors such as tapping or heartbeat (metaphor-based approach)~\cite{seifi2015vibviz}. 
After creating an initial Tacton set, the designer seeks to identify and select the most distinguishable (i.e., dissimilar) subset of Tactons for an application so that people can easily perceive the Tactons and learn their meanings~\cite{ryu2010vibrotactile}.

In contrast to extensive research and guidelines on the perceptual similarity of mechanical Tactons, limited guidelines exist on designing   distinguishable mid-air ultrasound Tactons. 
Decades of research on mechanical vibrations provide several temporal parameters for creating distinguishable Tactons such as amplitude, rhythm, and envelope frequency~\cite{hwang2017perceptual, kwon2023can, ternes2008designing, pasquero2006perceptual}. 
Yet, more research is needed on the efficacy of these temporal parameters for creating ultrasound Tactons.
Also, large sets of mechanical Tactons exist in the literature~\cite{ternes2008designing,van2003distilling,israr2014feel} and as open-source libraries~\cite{seifi2015vibviz}, but it remains unclear how these libraries can inform the design of ultrasound Tactons. 
To address these gaps, we ask: (1) Which temporal parameters can help create distinguishable mid-air ultrasound Tactons? 
Furthermore, to test if haptic designers can use their knowledge and existing resources on mechanical vibrations for ultrasound design, we ask a second question: 
(2) How does the distinguishability of ultrasound Tactons differ from the distinguishability of mechanical Tactons with corresponding temporal parameters?

To address these questions, we investigated the perceptual distinguishability of mid-air ultrasound Tactons that vary on temporal parameters in five user studies with 72 participants (Figure~\ref{fig:teaser}). The first study evaluated the discrimination threshold or Just Noticeable Differences (JND) for amplitude modulated frequencies (AM frequencies) as a temporal design parameter for mid-air ultrasound Tactons. The results for three AM frequencies (30\,Hz, 80\,Hz, and 210\,Hz) from 12 participants showed significant differences in JND values between 30\,Hz (JND = 47.2\%) and the other two reference AM frequencies (77.4\% for 80\,Hz, and 68.4\% for 210\,Hz).
Building on this data, we designed four sets of mid-air ultrasound Tactons based on existing mechanical Tactons and studied their perceptual distinguishability. 
Specifically, we controlled the temporal ultrasound parameters corresponding to the mechanical vibration parameters in three parameter-based Tacton sets~\cite{park2011perceptual, lim2023can, abou2022vibrotactile}. These Tactons varied in amplitude, envelope frequency, AM frequency, superposition ratio, and rhythm. 
For the fourth set, we selected a metaphor-based mechanical Tacton set from an open-source vibration library~\cite{seifi2015vibviz}, extracted their temporal envelopes and frequencies, and designed mid-air ultrasound Tactons with the corresponding temporal patterns.
We ran four user studies to collect pair-wise similarity ratings for the four sets of Tactons from 60 participants (n=15 participants per study).

We analyzed the perceptual dissimilarity spaces (i.e., distinguishability) of ultrasound Tactons from our studies and compared them to the perceptual spaces of mechanical Tactons in the literature. The analysis of mid-air ultrasound Tactons showed notable trends in user perception of the temporal parameters.
We found envelope frequency ($\leq$ 5\,Hz), rhythmic structure (i.e., the number and duration of pulses), and total duration can create distinguishable ultrasound Tactons.
Furthermore, our results revealed a strong correlation between similarity ratings for mid-air ultrasound and similarity of mechanical vibrations (mean Spearman's $\rho$=0.75 for Studies 2--5). 
This correspondence between the perceptual spaces of the two technologies suggests that the temporal parameters can provide distinguishable Tactons in both technologies. Notably, the Tacton set varied by rhythm showed the highest correlation (Spearman's $\rho$ = 0.89) between the two technologies, denoting rhythm as a key parameter for creating distinguishable Tactons in both technologies. Finally, we present differences in similarity ratings between mid-air ultrasound and mechanical Tactons, highlighting the distinct nature of the contact and contactless vibration technologies. 
For instance, the change in frequency spectrum contributed to the perceptual space for complex mechanical Tactons, but this parameter was not present in the perceptual space for the corresponding set of complex ultrasound Tactons. 
Based on the above studies, we present six design guidelines for creating distinguishable mid-air ultrasound Tactons with temporal parameters and discuss directions for future research on mid-air Tactons. 
Our contributions include:
\begin{itemize}
    \item JND values for AM frequency of mid-air ultrasound vibrations.
    \item Similarity ratings and perceptual spaces for four sets of mid-air ultrasound Tactons.
    \item Comparison of shared and distinct trends between distinguishability of mid-air ultrasound Tactons and distinguishability of mechanical vibrations.
     \item Six guidelines for designing mid-air ultrasound Tactons with temporal parameters for parameter-based and metaphor-based approaches.
\end{itemize}

\section{Related Work}
We review past research on designing mid-air ultrasound patterns followed by the design parameters of mechanical vibrations and studies of perceptual similarity in the haptics literature. 

\subsection{Design Parameters and Perception Studies for Mid-Air Ultrasound Haptics}
Mid-air ultrasound haptic technology can create diverse tactile sensations on human skin by modulating ultrasound signals using different techniques.
In the amplitude modulation (AM) technique, the amplitude of the ultrasound signal, with a carrier frequency above 20\,kHz, is modulated with a frequency below 1000\,Hz (i.e., AM frequency)~\cite{hoshi2009non, long2014rendering}.
With the spatiotemporal modulation (STM) technique, the focal point rapidly moves over any arbitrary path on the skin~\cite{frier2018using}. The moving speed of the focal point is known as drawing frequency. Others have combined these modulation techniques to improve user perception of ultrasound patterns. For example, Hajas et al. combined AM and STM techniques to create the feel of a slowly moving focal point on the skin. This technique improved user perception of different shapes rendered on the skin compared to using only STM~\cite{hajas2020mid}.

Previous research has investigated the thresholds of detecting and discriminating ultrasound patterns. 
Relevant studies have reported a U-shaped \textit{detection} threshold for various modulation techniques, with the minimum detection threshold at approximately 200\,Hz~\cite{takahashi2018lateral, hasegawa2018aerial, raza2019perceptually}.
Furthermore, Hasegawa and Shinoda found that the \textit{detection} threshold increased between 300\,Hz to 1000\,Hz for the AM technique~\cite{hasegawa2018aerial}.
Howard et al. estimated \textit{detection} thresholds of ultrasound amplitude with the AM and STM techniques~\cite{howard2019investigating}.
They found that the static focal point with 200\,Hz AM frequency showed a higher amplitude \textit{detection} threshold of 556.9\,Pa compared to the 334.1\,Pa \textit{detection} threshold by the STM moving the focal point along a line path. 
Regarding \textit{discrimination} thresholds or just noticeable differences, prior research has investigated the JND for the STM drawing frequency~\cite{rutten2020discriminating, wojna2023exploration}.
Wojna et al. showed an average JND of 20.7\%\ for five reference drawing frequencies ranging from 30\,Hz to 70\,Hz~\cite{wojna2023exploration}.
Also, for six reference drawing frequencies ranging from 1\,Hz to 2\,Hz with the step size of 0.2\,Hz, the average JND was 25.5\%\ when the moving focal point vibrated at a 125\,Hz AM frequency~\cite{rutten2020discriminating}.
Our work complements this literature by contributing data on the JND for AM frequency as another temporal parameter for mid-air ultrasound haptic design.

Other researchers investigated user perception and experience of various ultrasound Tactons. 
Specifically, past research has proposed rendering algorithms for two or three-dimensional tactile shapes~\cite{long2014rendering, korres2016haptogram} and evaluated the identification accuracy of the mid-air ultrasound shapes through user studies~\cite{long2014rendering, korres2016haptogram, rutten2019invisible, hajas2020mid}. 
Obrist et al. reported how users described the feel of two AM frequencies (16\,Hz and 250\,Hz) and linked the descriptions to the activation of two types of mechanoreceptors, Meissner and Pacinian corpuscles~\cite{obrist2013talking}. 
Others investigated emotional ratings and user descriptions for various design parameters of ultrasound Tactons~\cite{obrist2015emotions,dalsgaard2022user}.
Recent work has investigated ultrasound Tacton design for automotive applications~\cite{brown2020ultrahapticons, brown2022augmenting}.
These studies followed a top-down use case-driven approach to ultrasound Tacton design, focusing on the identification of ultrasound Tactons with visual or textual references. In contrast, 
our work contributes data on the distinguishability of mid-air ultrasound Tactons varying on temporal parameters using pairwise comparisons and no visual stimuli to identify perceptual patterns regardless of end applications.

\subsection{Design Parameters and Libraries for Mechanical Vibrations}
Haptic researchers have proposed many design parameters for mechanical vibrations in recent decades. Several studies vary parameters of a sinusoidal vibration, such as amplitude~\cite{ryu2010psychophysical, hwang2013vibrotactile}, frequency~\cite{hwang2010perceptual, israr2006frequency, tan1999information}, envelope modulation frequency~\cite{park2011perceptual}, duration~\cite{yoo2015emotional, kwon2023can}, superposition ratio between two sinusoids~\cite{hwang2017perceptual, lee2013real, yoo2022perceived}, and mixtures of multiple parameters~\cite{lim2023can, yoo2015emotional}.
Inspired by musical notes and structure, other researchers proposed rhythm, pulse or note length~\cite{ternes2008designing, abou2022vibrotactile}, and sound waveform (or timbre)~\cite{brewster2004tactons, brown2005first, pasquero2006perceptual} as new parameters for mechanical vibrations. Haptic designers can create mechanical Tactons that systematically vary on one or more of these parameters to study their impact on user perception~\cite{park2011perceptual} or emotion~\cite{yoo2015emotional,seifi2013first}. We select three Tacton sets that cover the above parameters, except for duration and waveform.

Besides the parameter-based approach, designers can create new Tactons by modifying existing examples from a Tacton library~\cite{schneider2016studying}. Haptic researchers have proposed various Tacton libraries for mechanical Tactons. Van Erp and Spap\'e designed 59 Tactons by transforming short musical pieces to vibrations~\cite{van2003distilling}. Haptic Muse by Immersion Inc. included 124 Tactons for different game effects~\cite{immersionmuse}.  FeelEffects by Disney Research was a library of over 50 vibrations for a haptic seat~\cite{israr2014feel}. VibViz was developed as a library of 120 mechanical vibration Tactons annotated with user ratings and descriptive tags on the vibrations' physical, sensory, emotional, usage, and metaphoric attributes~\cite{seifi2015vibviz}. These libraries included Tactons with different durations, complex temporal envelopes, and wide frequency spectrums. We used 14 mechanical Tactons from VibViz, the only open-source vibration library, and investigated the distinguishability of the corresponding ultrasound Tactons with complex temporal patterns.

To inform the design of mechanical Tactons, past studies have provided data on the discrimination thresholds for the frequency of mechanical vibrations~\cite{goff1967differential, franzen1975vibrotactile, rothenberg1977vibrotactile, goble1994vibrotactile, israr2006frequency}. 
Choi and Kuchenbecker aggregated these findings and concluded that the JND values for the frequency of mechanical vibrations mostly fall between 15\% and 30\%~\cite{choi2012vibrotactile}. 
We compare the frequency JND for mechanical vibrations with our data for mid-air ultrasound vibrations to discuss differences in the distinguishability of Tactons between the two haptic technologies.

\subsection{Similarity Perception for Tactons}

The perceptual similarity of natural and programmable haptic stimuli has been an important research topic in haptics. 
To derive the perceptual space for a Tacton set, designers run a user study to collect similarity ratings between Tacton pairs in the set, then apply multi-dimensional scaling (MDS) to obtain a low-dimensional visualization of the perceptual distances of the Tactons. The designers then analyze this perceptual space to identify the dominant parameters that can describe the spatial configuration of the Tacton similarities.
Several studies examined the perceptual space of the mechanical vibration Tactons~\cite{hwang2010perceptual, hwang2017perceptual, park2011perceptual, abou2022vibrotactile, lim2023can, ternes2008designing, pasquero2006perceptual}.
Specifically, some studies examined the perceptual space of rhythm and sound waveform~\cite{pasquero2006perceptual, ternes2008designing, kwon2023can}.
Hwang et al. showed that the superimposed vibration formed a half-circular shape outside the linear distance between the two sinusoids, suggesting the superposition ratio could be a distinct design parameter for mechanical vibrations~\cite{hwang2017perceptual}.
Park et al. studied envelope frequencies ranging from 0\,Hz to 80\,Hz and found that lower envelope frequencies ($\leq$ 20\,Hz) can contribute to distinguishable Tactons~\cite{park2011perceptual}.
Others explored the perceptual spaces of vibrations that vary on multiple parameters simultaneously and identified the subset of dominant parameters such as envelope frequency and rhythm~\cite{ternes2008designing,lim2023can}.
These studies inform haptic design by highlighting the efficacy of various temporal parameters for creating distinguishable mechanical Tactons. 
We follow the same study procedure to analyze the perceptual space of mid-air ultrasound Tactons with temporal parameters.

Past studies also established methods for evaluating correspondence between two perceptual spaces.
Vardar et al. collected pair-wise similarity ratings for ten natural textures in the visual and haptic modalities and showed high correspondence between the perceptual spaces in the two modalities using Spearman rank correlation of the similarity ratings~\cite{vardar2019fingertip}.
Others showed high Spearman rank correlation and consistent perceptual spaces for mechanical Tactons in the lab and crowdsourced settings~\cite{kwon2023can,abou2022vibrotactile}.
Similar to these studies, we use the Spearman rank correlation and visualization of the perceptual spaces to compare the similarity ratings for ultrasound Tactons with results for mechanical Tactons in the literature.

\section{Approach}
\begin{figure*}[t]

    \centering
  
    \begin{subfigure}[t]{1.0\textwidth}
    \includegraphics[width=\linewidth]{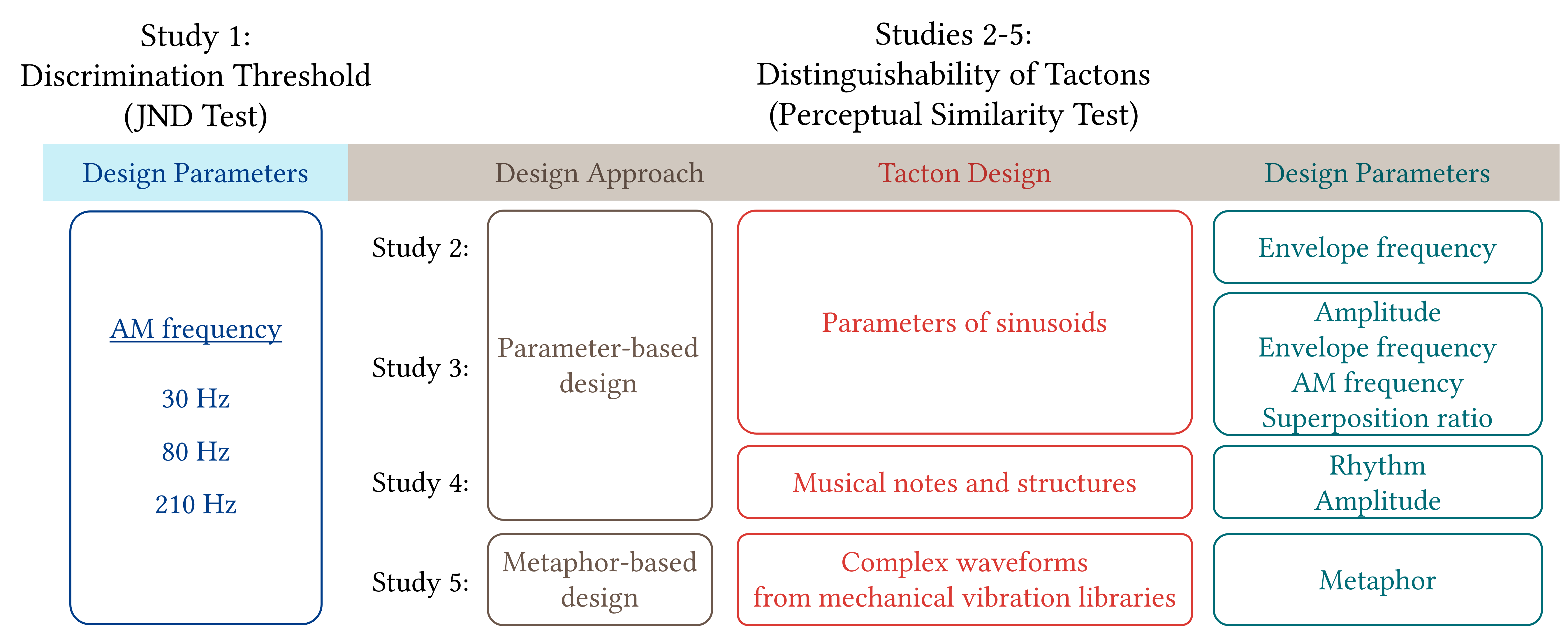}
    \end{subfigure}
    
    \caption{ An overview diagram for Studies 1--5.
    In Study 1, we examined the discrimination thresholds for AM frequencies of 30\,Hz, 80\,Hz, and 210\,Hz.
    In Studies 2--5, we investigated the perceptual distinguishability for four sets of mid-air ultrasound Tactons that vary in their temporal parameters.
    We selected the four sets of mechanical Tactons from~\cite{park2011perceptual, lim2023can, ternes2008designing, seifi2015vibviz} to create the ultrasound Tactons.
    }
    \label{fig:studyoverview}
    \Description{This diagram provides an overview of the five studies presented in this paper. Study 1 examines the discrimination threshold of amplitude-modulated frequencies at 30Hz, 80Hz, and 210Hz to evaluate their viability as a design parameter. Studies 2-5 explore the perceptual distinguishability of four ultrasound Tacton sets, each varying in design approach, Tacton design, and specific parameters. The Tacton sets in Studies 2-4 are created using a parameter-based approach, while the set in Study 5 employed a metaphor-based approach. In Studies 2-3, Tactons are designed with parameters of sinusoids; in Study 4, they are designed with musical notes and structures; and in Study 5, they are designed with complex waveforms from mechanical vibration libraries. Regarding design parameters, the Tacton set in Study 2 varies envelope frequency; the set in Study 3 varies amplitude, envelope frequency, amplitude-modulated frequency, and superposition ratio; the set in Study 4 varies rhythm and amplitude; the set in Study 5 varies metaphor.}

\end{figure*}

We conducted five studies to assess the efficacy of temporal parameters for ultrasound Tacton design (Figure \ref{fig:studyoverview}). We followed consistent experimental settings in all the studies as described below.
The studies were approved by the Institutional Review Board at Arizona State University (STUDY00017208).

\subsection{Selection of Temporal Parameters for Mid-Air Ultrasound Tactons}
\label{sec:TactonSetSelection}
In Study 1, we investigated the JND for AM frequency of mid-air ultrasound vibrations (Section~\ref{sec:UserStudy1}). 
Next, we created four sets of mid-air ultrasound Tactons and estimated their perceptual dissimilarities based on user pair-wise ratings in Studies 2--5 (Section~\ref{sec:Studies2-3}). 
We selected these four sets from the mechanical Tacton literature that reported the similarity rating data and the perceptual spaces (Figure~\ref{fig:studyoverview}).
We selected two sets that varied in parameters of sinusoids~\cite{park2011perceptual, lim2023can}.
The first set included eight Tactons varying on their envelope frequencies, and the second set had 12 Tactons with combinations of four parameters: amplitude, envelope frequency, frequency, and superposition ratio.
The third set had 14 Tactons varying on seven rhythms and two amplitudes~\cite{abou2022vibrotactile}.
The three Tacton sets covered most of the temporal design parameters for mechanical Tactons.
We did not study Tactons that varied in waveform shape (e.g., square waves) since mid-air ultrasound technology does not offer this design parameter.

We selected the fourth Tacton set based on 14 mechanical vibrations that were designed to remind people of metaphors, with two Tactons per seven metaphors: \textbf{Alarm}, \textbf{Animal}, \textbf{Engine}, \textbf{Heartbeat}, \textbf{Jumping}, \textbf{Tapping}, and \textbf{Walking}~\cite{kwon2023can}.
These Tactons were from a large mechanical Tacton library (i.e., VibViz)~\cite{seifi2015vibviz}, had more complex temporal patterns than the previous sets, and their durations ranged from 0.84 to 5.38 seconds (Figure~\ref{fig:Signal_UserStudy5}).
Therefore, our investigation could provide insights into creating mid-air Tactons using existing libraries for mechanical Tactons.

\subsection{Constant Spatiotemporal Stimulation Pattern}
\label{sec:ultra-stimuli}
Since this work focused on the temporal parameters of ultrasonic mid-air haptics, we kept the spatiotemporal parameters constant as follows.
We used STM to render a circular trajectory in all five studies, while the focal point vibrated by AM (Figure~\ref{fig:JND_hand}).
This approach was previously used in the ultrasonic mid-air haptics literature to provide a stronger tactile sensation than rendering a single static focal point~\cite{hajas2020mid, rutten2020discriminating}.
Also, we used the single-point STM as this technique offers higher perceived intensity than multiple-point STM~\cite{shen2023multi}.
Considering users' typical palm size, we rendered a circle with a radius of 10\,$mm$ and a drawing speed of 12\,$m/s$ at 20\,$cm$ above the device, following previous work~\cite{dalsgaard2022user, raza2019perceptually, frier2019sampling}.
These choices aimed to ensure that all ultrasound Tactons in Studies 1-5 were perceivable by users while preventing users from perceiving the shape and the movement of the focal point trajectory.
The diameter for the focal point's trajectory was close to the spatial resolution of a single focal point ($\sim$20\,$mm$~\cite{wilson2014perception}), and the drawing frequency exceeded the motion perception thresholds~\cite{freeman2021perception}.
In a pilot test, three users freely tested various STM parameters and AM frequencies and confirmed that the aforementioned STM parameters (10\,$mm$, 12\,$m/s$) provided salient stimulation without perceiving a motion. 
In Section~\ref{sec:preliminary}, we examined the generalizability of our studies by testing other STM combinations of radius and drawing speed.

\subsection{Experiment Setup}
We used the STRATOS Explore device by Ultraleap~\cite{stratos} to render the mid-air ultrasound vibrations.
The device provides an intensity range between 0\%\ and 100\%, which corresponds to the peak acoustic radiation pressure at the focal point, ranging from 0 to 1125\,Pa~\cite{stratos}.
The device was on a table in front of the participants during the studies. We used an armrest to ensure the participant's palm was 20\,$cm$ above the device's center.
We collected the participant responses with a graphical user interface (GUI) on an iOS mobile phone (Study 1) and a Desktop computer (Studies 2--5).
Participants used their dominant hand to interact with the GUI programs and felt the ultrasound patterns on their non-dominant hand.
Also, the participants wore noise-canceling headphones with white noise to block any sound from the device.
Throughout the studies, we maintained the room temperature between 20--25 degrees Celsius.

\section{Study 1: Just Noticeable Difference for AM Frequency of Mid-Air Ultrasound Vibrations}
\label{sec:UserStudy1}

\begin{figure*}[t]
  \centering

  \begin{subfigure}[t]{0.3\textwidth}
  \centering
    \includegraphics[height=4cm]{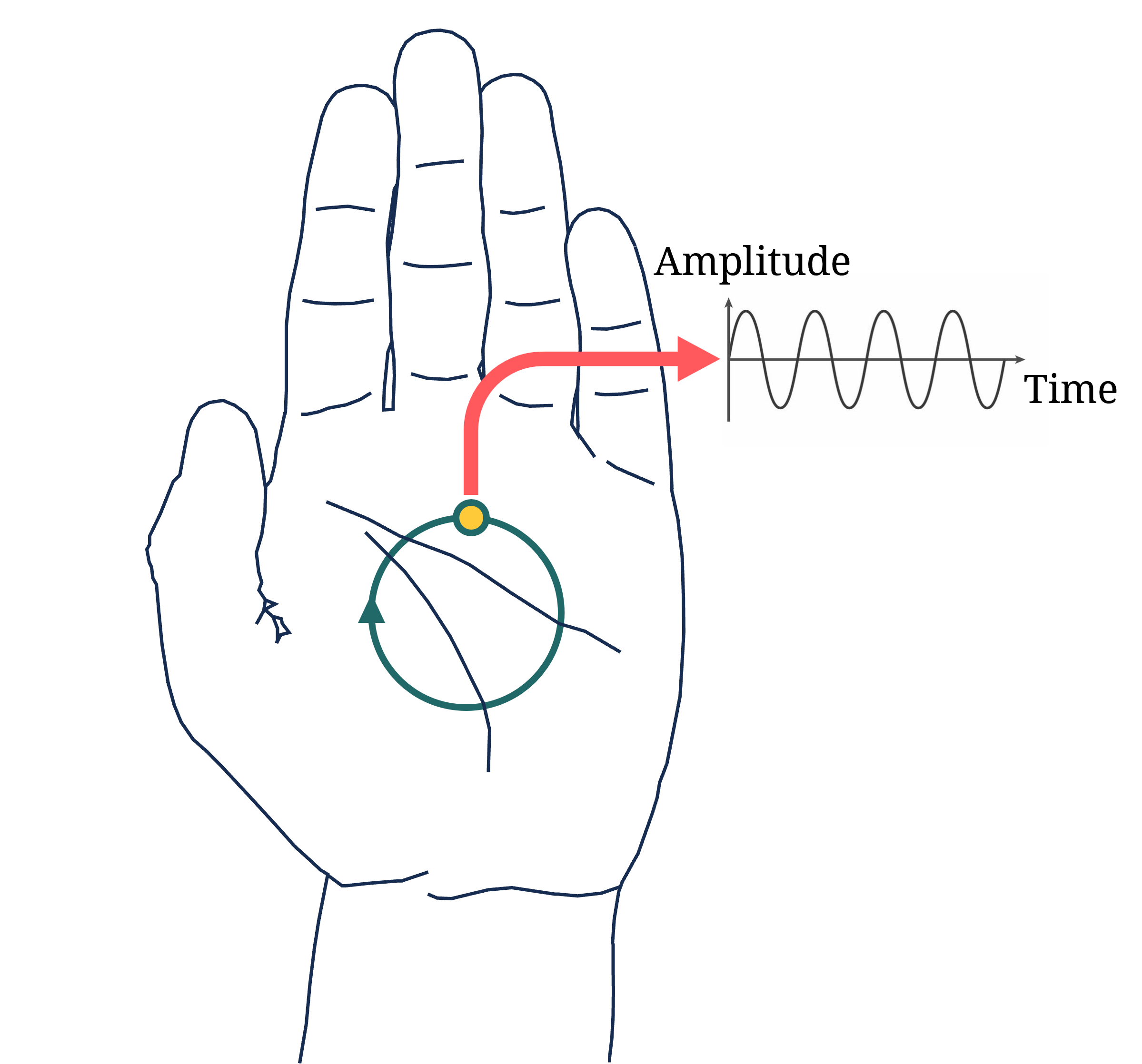}
    \caption{}
    \label{fig:JND_hand}
  \end{subfigure}
  \hfill
  \begin{subfigure}[t]{0.3\textwidth}
  \centering
    \includegraphics[height=4cm]{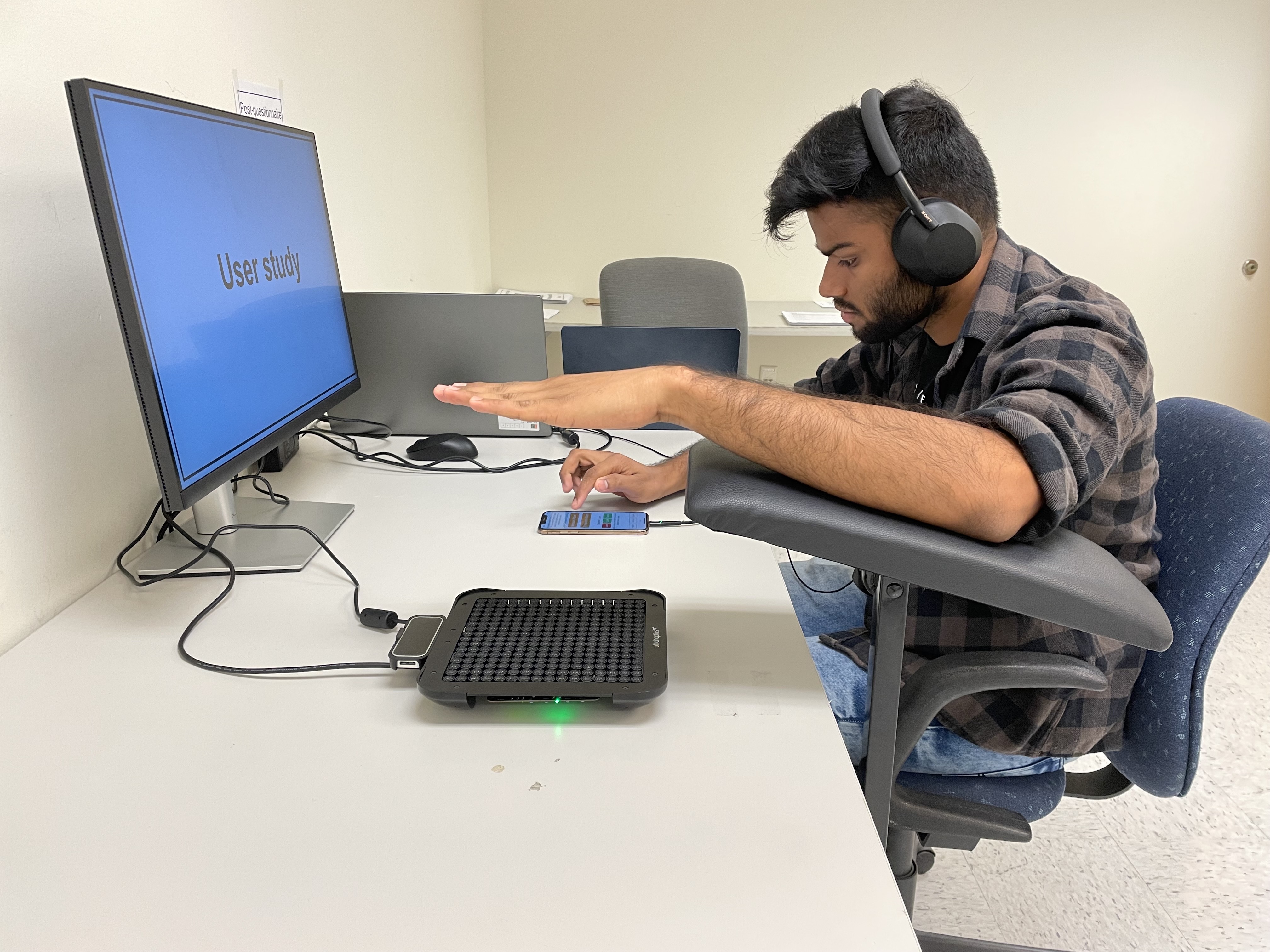}
    \caption{}
    \label{fig:JND_setup}
  \end{subfigure}
  \hfill
  \begin{subfigure}[t]{0.3\textwidth}
  \centering
    \includegraphics[height=4cm]{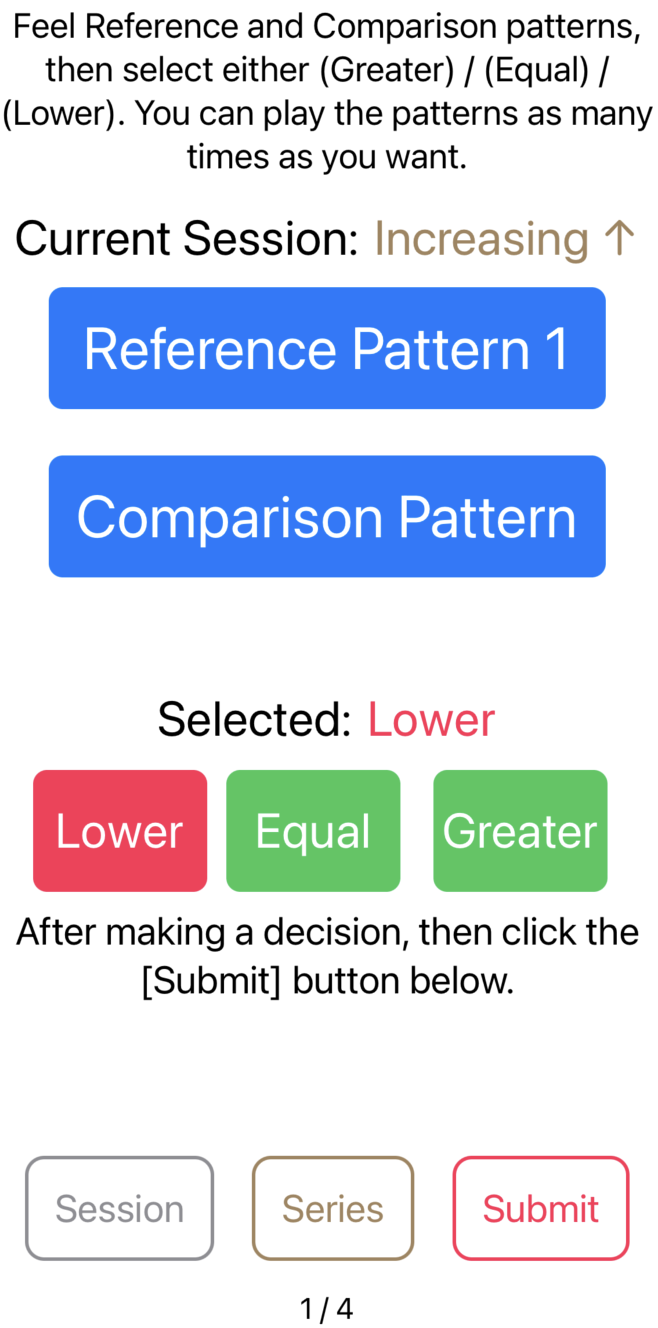}
    \caption{}
    \label{fig:JND_GUI}
  \end{subfigure}

  \caption{Stimuli and set up for the studies: (a) The spatiotemporal stimulation used for Studies 1--5. We maintained a radius of 10~$mm$ and a drawing speed of 12\,$m/s$ with a single focal point STM, (b) Experimental setup used for Study 1, and (c) A screenshot of the GUI program used to collect user responses in Study 1. The selected response was highlighted in red. The color did not indicate the correctness of the response.}
  \label{fig:JND_Experiment}
  \Description{Three photos labeled (a), (b), and (c) show the configuration of spatiotemporal stimulation in Studies 1-5, the experimental setup of Study 1, and a screenshot of the smartphone GUI program used in Study 1, respectively. Image (a) shows a palm with a moving focal point following a circular trajectory with a 10mm radius. The focal point vibrates using the amplitude modulation technique. Image (b) shows a participant seated in a chair, holding a smartphone in the right hand and placing the left hand on the armrest, perceiving ultrasound vibrations from a distance. Image (c) shows two buttons at the top, allowing the user to feel a reference pattern and a comparison pattern; three buttons in the middle labeled “Lower,” “Equal,” and “Greater” for assessing the two patterns; and three buttons at the bottom labeled “Session,” “Series,” and “Submit,” related to the experimental process.}
\end{figure*}

We investigated the discrimination thresholds or just noticeable differences (JND) for AM frequencies to inform the design of distinguishable mid-air ultrasound Tactons.

\begin{figure*}[t]
  \centering

  \begin{subfigure}[c]{0.34\textwidth}
    \includegraphics[width=\linewidth]{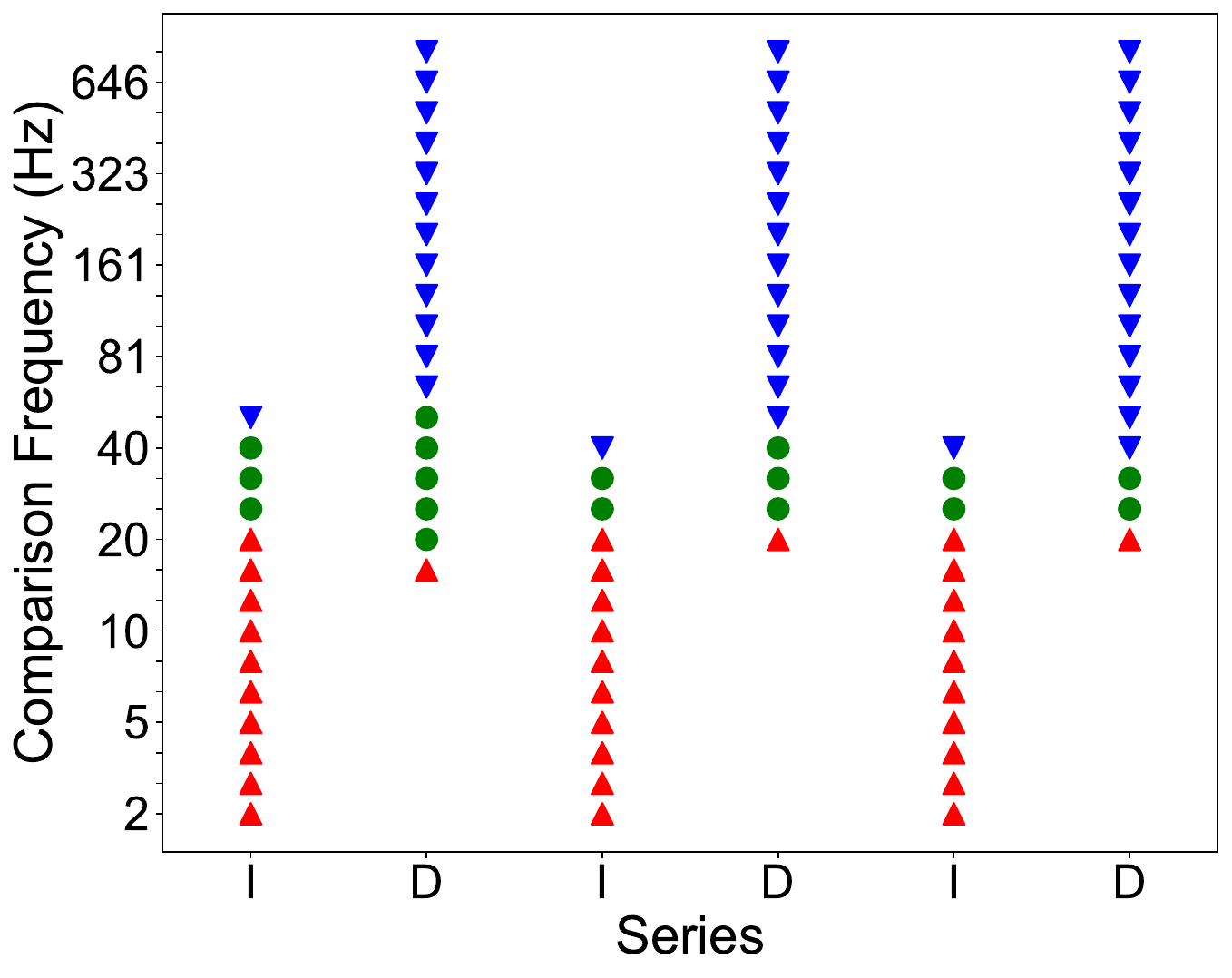}
  \end{subfigure}
  \hfill
  \begin{subfigure}[c]{0.34\textwidth}
    \includegraphics[width=\linewidth]{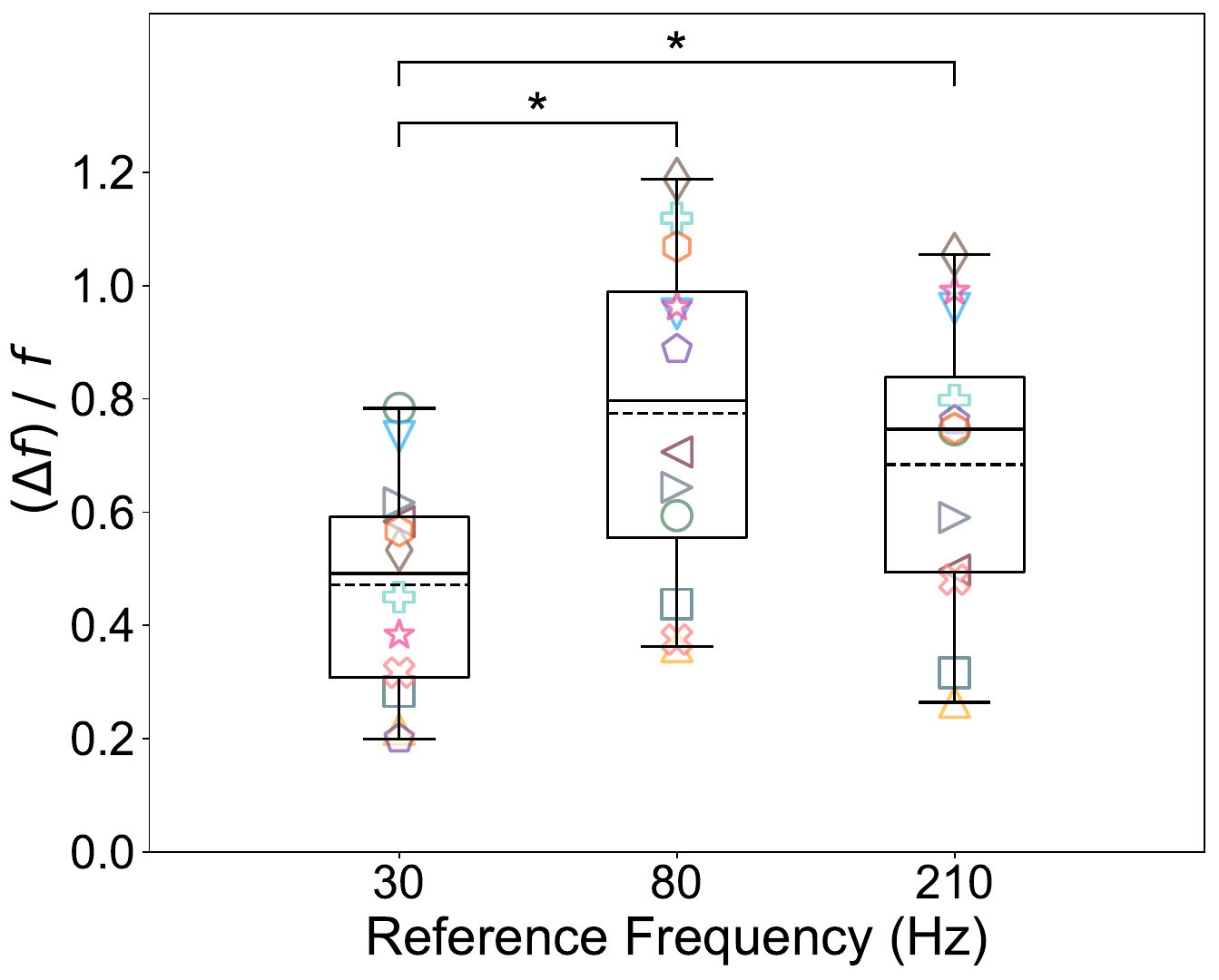} 
  \end{subfigure}
  \hfill
    \begin{subfigure}[c]{0.29\textwidth}
    \includegraphics[width=\linewidth]{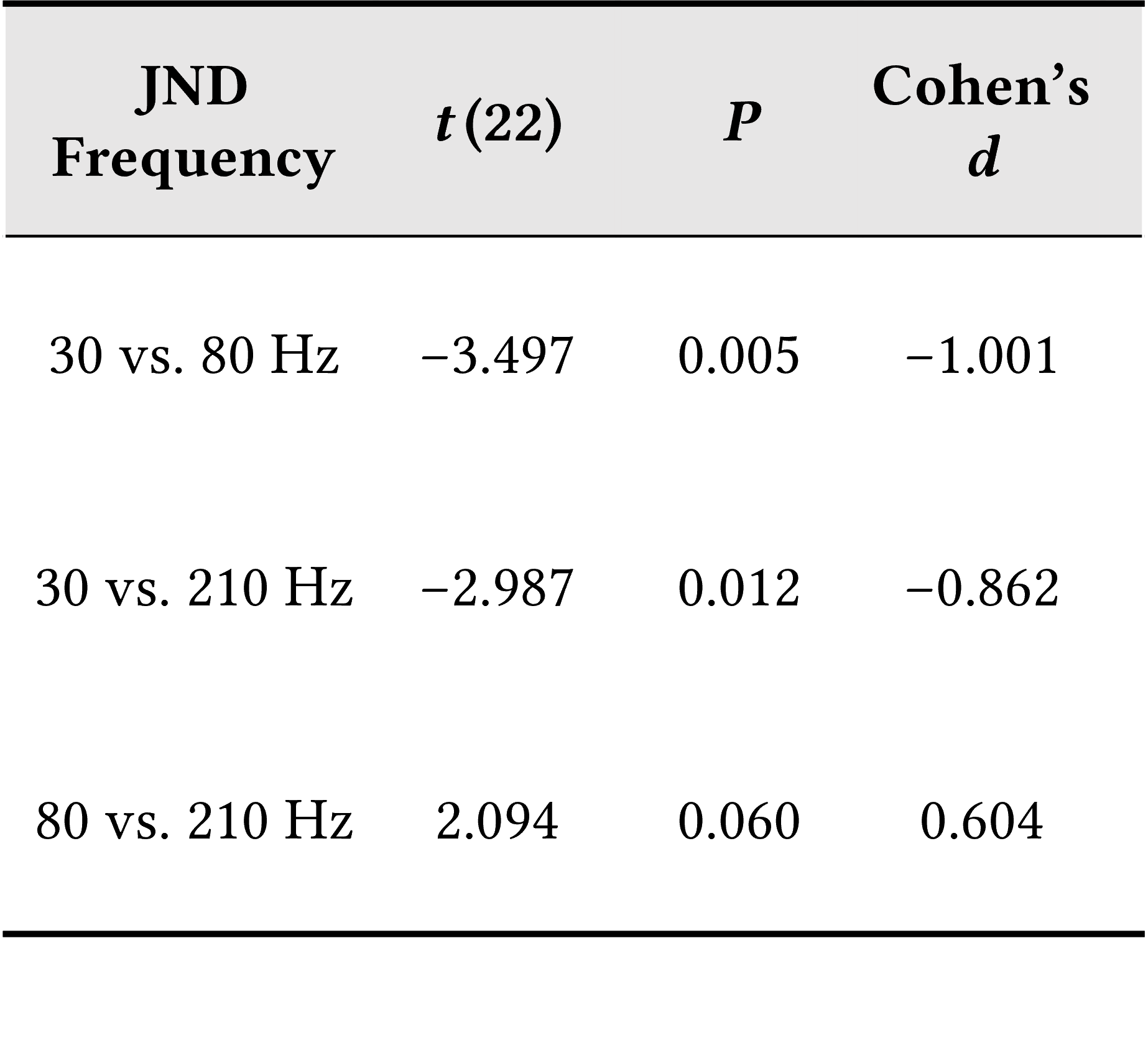} 
  \end{subfigure}

    \begin{subfigure}[t]{0.34\textwidth}
    \caption{}
    \label{fig:JND_Example}
    \end{subfigure}
    \hfill
    \begin{subfigure}[t]{0.34\textwidth}
    \caption{}
    \label{fig:JND_Plot}
    \end{subfigure}
    \hfill
    \begin{subfigure}[t]{0.29\textwidth}
    \caption{}
    \label{fig:post-hoc}
    \end{subfigure}
    
  \caption{ Results of Study 1 on the Just Noticeable Differences (JND) for AM frequencies of ultrasonic mid-air stimuli with 12 participants. (a) An example of a study session with the reference AM frequency of 30\,Hz. ``I'' represents an increasing series of frequencies, and ``D'' a decreasing series. Red, green, and blue colors denote ``lower'', ``equal'', and ``greater'' responses from the participant, respectively.
  (b) JNDs for AM frequency for each participant and average values. An asterisk (*) indicates a significant difference in JNDs. The dotted and solid lines denote the mean and median, respectively.
  (c) Post-hoc paired t-tests for JND values.
  }
  \Description{Three plots labeled (a), (b), and (c) show an example plot of a session of user study and the results in Study 1. Plot (a) shows a session that alternates three increasing series and three decreasing series, when reference frequency is 30Hz. Starting from 2Hz at increasing series or 814Hz at decreasing series, the comparison frequencies change in the direction of the reference frequency, successively. And a user compares the comparison frequency to the reference frequency, 30Hz, and rate either “lower”, “equal”, or “greater”. Each rating was graphically illustrated by a red triangle for the “lower”, a green circle for the “equal”, and a blue triangle for the “greater”. In the increasing series, the series ends once the user reports “greater”. In contrast, in the decreasing series, the series ends once the user reports “lower”. Plot (b) shows all participants’ just noticeable difference results for amplitude-modulated frequencies at 30Hz, 80Hz, and 210Hz. Average values are 47.2\%, 77.4\%, and 68.4\% at 30 Hz, 80 Hz, and 210 Hz, respectively. And corresponding standard deviations are 18.7\%, 28.1\%, and 24.7\%. The significant differences are observed between 30Hz and 80Hz and between 30Hz and 210Hz. Plot (c) shows post-hoc paired t-tests for JND values. For JND values of 30Hz and 80Hz, t(22) = -3.497, p = 0.005, Cohen's d = -1.001; For JND values of 30Hz and 210Hz, t(22) = -2.987, p = 0.012, Cohen's d = -0.862; For JND values of 80Hz and 210 Hz, t(22) = 2.094, p = 0.060, Cohen's d = 0.604.}
    
\end{figure*}

\subsection{Method}
\subsubsection{Stimuli}
We selected three reference AM frequencies of 30\,Hz, 80\,Hz, and 210\,Hz for this study.
These frequencies cover the ranges commonly used in previous experiments for both mid-air ultrasound and mechanical vibrations~\cite{lim2023can, lee2013real, hwang2010perceptual, raza2019perceptually, dalsgaard2022user, obrist2015emotions}.
Also, this choice ensured that the participants could perceive the differences among the reference AM frequencies based on a pilot study and helped keep the study session under 2 hours.
We fixed the amplitude of all stimuli at the maximum (100\%) on the mid-air ultrasound device across all AM frequencies, similar to the past JND studies for drawing frequency in ultrasonic mid-air haptics~\cite{rutten2020discriminating, wojna2023exploration}.

\subsubsection{Participants}
We recruited 12 participants (three females and nine males, 20--27 years old) by advertising the study on the university mailing lists.
The participants were all right-handed and had no sensory impairments in their hands. 
On average, the study took 88 minutes per user, and the participants received a \$30 USD Amazon gift card.

\subsubsection{Experiment Procedure}
To investigate JND for AM frequency, we chose the method of limits due to its efficiency and reliable results \cite{gescheider2013psychophysics}.
In this method, the participant feels a reference and comparison stimuli in pairs and selects whether the comparison stimulus has a lower, equal, or higher value than the reference stimulus.
For our study, we prepared 26 comparison stimuli ranging from 2\,Hz to 814\,Hz, using an exponential sequence of one-third-octave frequency intervals (26\%), based on a previous JND study for the frequency of mechanical vibrations~\cite{israr2006frequency}.
All stimuli lasted for 1 second, with a gap of at least 1 second between the reference and comparison stimuli pairs.
We collected user responses using the GUI program in Figure~\ref{fig:JND_GUI}.

First, the participants answered a pre-questionnaire about their age, gender, and dominant hand and completed a practice session.  
The first session was a practice with a reference AM frequency of 130\,Hz and consisted of two series: one increasing series starting from 2\,Hz and one decreasing series starting from 814\,Hz.
Participants were unaware that this session was for practice. 
We discarded the results from the analysis.

The subsequent three sessions used the three reference stimuli. We determined the presentation order with a balanced Latin square.
Each session included six series, alternating between increasing and decreasing series. 
In each series, the pre-defined frequencies of the comparison stimuli successively increased starting from the lowest frequency (2\,Hz) in the increasing series or decreased from the highest frequency (814\,Hz) in the decreasing series.
For each stimuli pair, the participants experienced both the reference and comparison stimuli and responded whether the comparison stimulus had a frequency ``lower'', ``equal'', or ``greater'' than the reference stimulus.
Participants could select only one response.
Each series ended when the participants submitted a ``greater'' response in the increasing series and a ``lower'' response in the decreasing series (Figure~\ref{fig:JND_Example}). 
The participants could repeatedly play the stimuli or take breaks as necessary.
The study ended by completing a post-questionnaire about any discomfort experienced during the experiment and whether the stimuli were difficult to distinguish.

\subsection{Results}
We calculated the JNDs for each reference AM frequency by dividing the differential thresholds from the method of limits with the reference AM frequency~\cite{gescheider2013psychophysics}.
On average, the JNDs were 47.2\%, 77.4\%, and 68.4\% at 30\,Hz, 80\,Hz, and 210\,Hz, respectively, with standard deviations of 18.7\%, 28.1\%, and 24.7\% across the participants.
The JND values at each reference AM frequency met the assumptions of normality and homogeneity of variance.
A one-way repeated measures ANOVA test showed a statistically significant difference in the JND values for the three reference frequencies ($F(2, 22) = 10.051$; $p<0.001$; Cohen's $f = 0.497$).
Post-hoc paired t-tests showed a significant difference ($\alpha < 0.05$) in JND values between 30\,Hz and 80\,Hz as well as between 30\,Hz and 210\,Hz. 
We did not find a significant difference in JND values between 80\,Hz and 210\,Hz (Figure~\ref{fig:post-hoc}).

In the post-questionnaire, five participants reported experiencing difficulty in distinguishing higher frequencies.
We did not find any significant differences in the JND results based on gender, completion time, or the presentation order of the reference frequencies.

\subsection{Discussion}

The significant difference in JNDs of 30\,Hz and 80\,Hz can be due to the differences in the activation of mechanoreceptors in the skin, specifically Meissner and Pacinian corpuscles, in these frequency ranges. Past studies showed that frequencies below 40\,Hz primarily activated the Meissner corpuscles, while frequencies above 64\,Hz mostly activated the Pacinian corpuscles~\cite{johansson1982responses,gescheider2001frequency}.
The variations in mechanoreceptor activation by the frequencies may influence the differences in JNDs between 30\,Hz and 80\,Hz.
Also, previous studies showed that the amplitude detection thresholds increase for frequencies higher than 300\,Hz in mid-air ultrasound perception~\cite{hasegawa2018aerial}. 
Thus, the perceived intensity in the higher frequency ranges of comparison stimuli (323\,Hz to 814\,Hz) could have decreased when keeping the physical amplitudes constant in our study, leading to significantly higher JND results for 80\,Hz and 210\,Hz.

The significant difference between JNDs of  30\,Hz and 80\,Hz for AM frequencies in this study aligns with the significant difference reported between JNDs of 30\,Hz and 70\,Hz for the drawing frequency (i.e., speed of moving the focal point along a path)~\cite{wojna2023exploration}.
While AM frequency and drawing frequency are fundamentally different, these parallel results may hint at a change in ultrasound perception at approximately 70--80\,Hz regardless of the mechanisms used to generate frequency in ultrasonic mid-air haptics.
Our JND results for AM frequency (47\%--77\%) are higher than the results of previous JND studies for drawing frequency (20\%--25\%)~\cite{rutten2020discriminating, wojna2023exploration}, perhaps due to the lower perceived intensity of the ultrasound with AM compared to the STM technique.
Also, our AM frequency JND results were higher than the frequency JND for mechanical vibrations (around 15\%--30\%)~\cite{choi2012vibrotactile, israr2006frequency}, which may be due to the lower perceived intensity of mid-air ultrasound haptic stimuli and the contactless nature of the technology.

\begin{figure*}[t]
  \centering
  
  \begin{subfigure}[t]{0.48\textwidth}
  \centering
    \includegraphics[width=\linewidth]{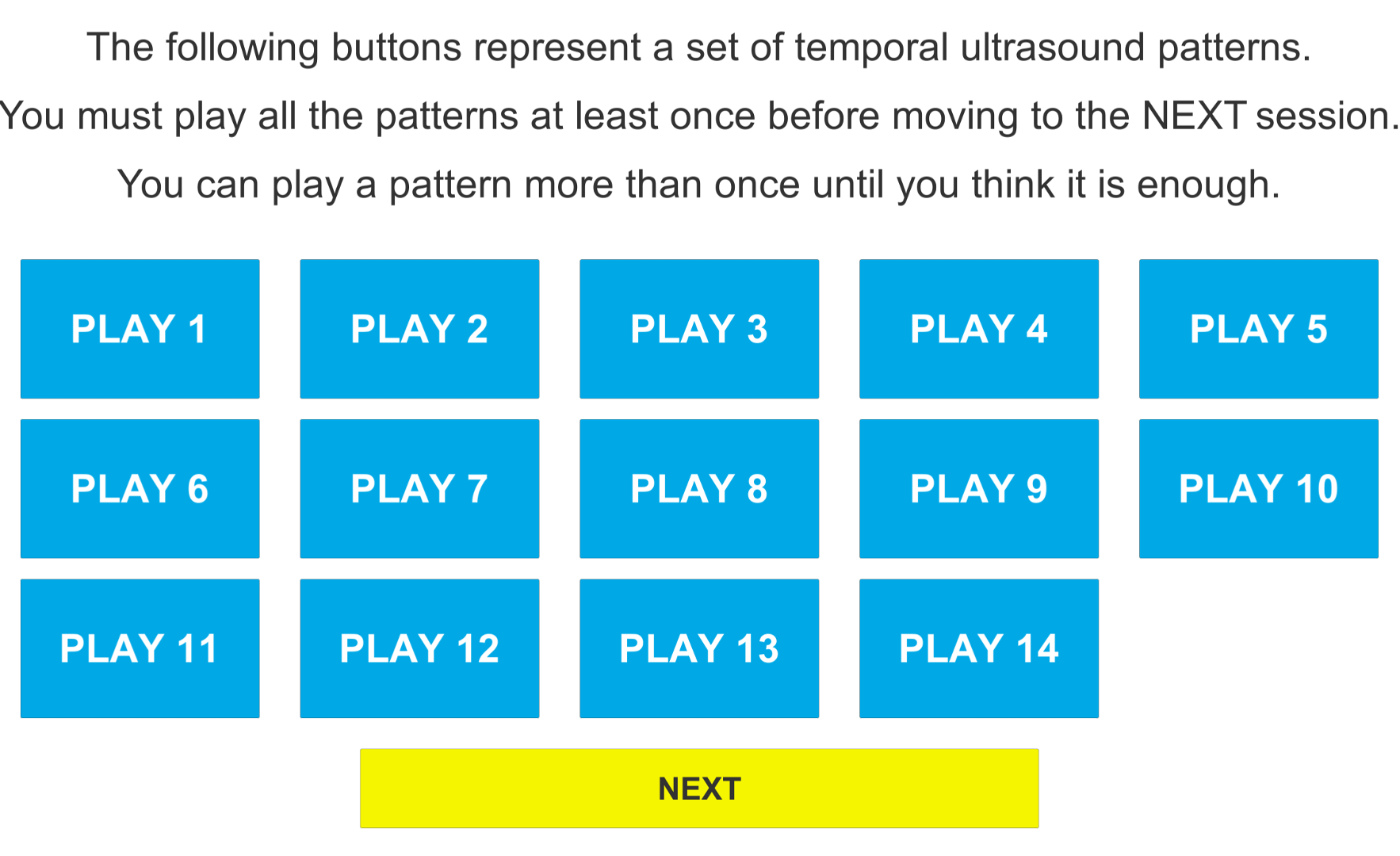}
    \caption{}
    \label{fig:rating_training}
  \end{subfigure}
  \hfill
  \begin{subfigure}[t]{0.48\textwidth}
  \centering
    \includegraphics[width=\linewidth]{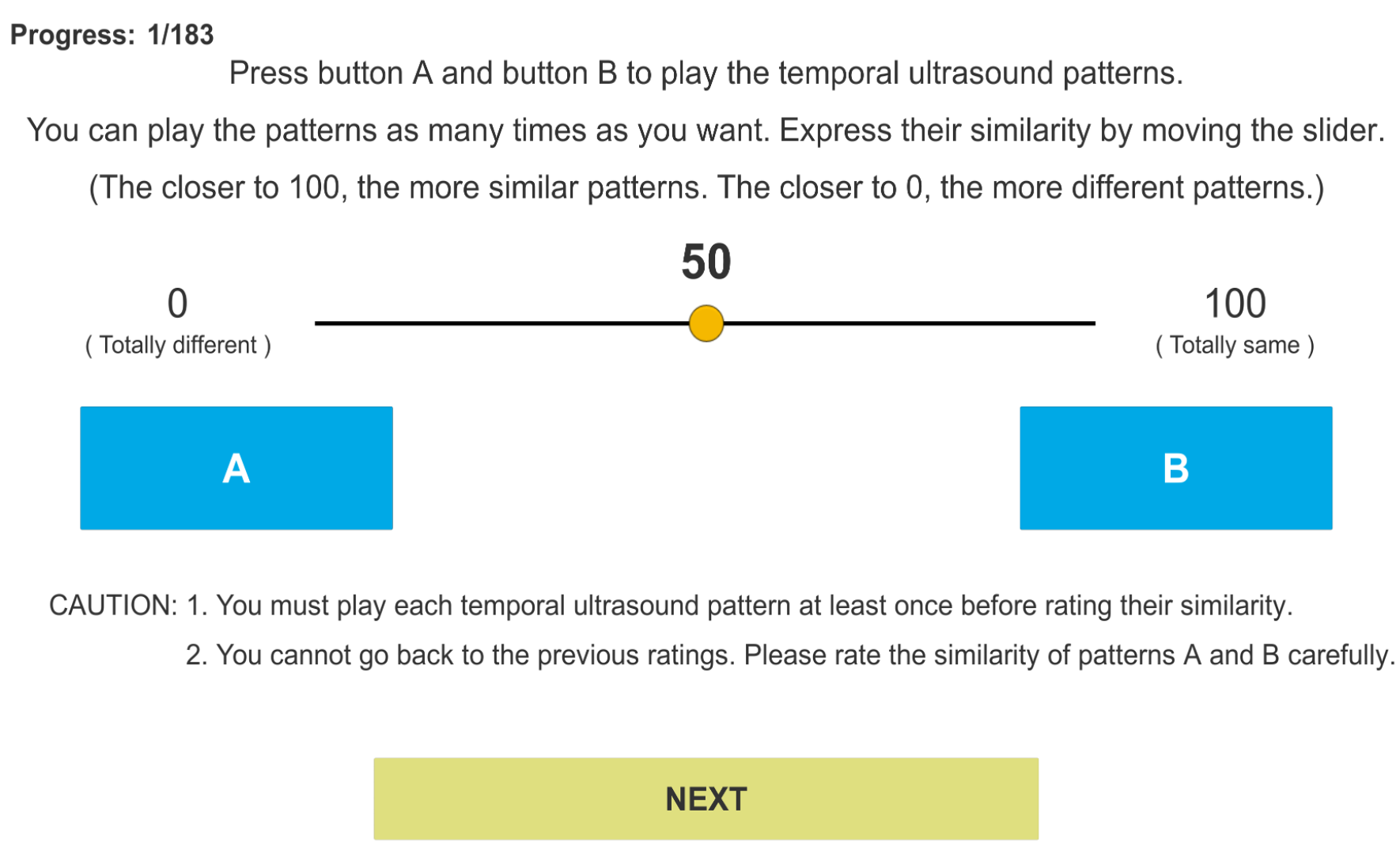}
    \caption{}
    \label{fig:rating_main}
  \end{subfigure}

  \caption{Screenshots of the graphical user interface for the (a) training and (b) main pair-wise similarity rating sessions in Studies 2--5.}
  \label{fig:rating_Experiment}
  \Description{Two images labeled (a) and (b) show screenshots of the GUI program used in Studies 2-5 to collect users’ perceptual similarity ratings. Image (a) illustrates a training session in the experiment. Users experience all ultrasound Tactons in a set by clicking buttons. Then, the “NEXT” button becomes active, guiding them to the main session. Image (b) illustrates the main session, where users experience Tacton pairs and rate their perceived similarity by adjusting a slider bar. This process is repeated throughout the main session.}
\end{figure*}

\textbf{Design Guidelines:} Haptic designers should consider the JND values when using AM frequency as a design parameter for mid-air ultrasound Tactons. 
Specifically, the designers should use a larger step size to AM frequencies above 80\,Hz for designing Tactons with drawing frequencies or for ensuring distinct sensations as the mechanical vibrations in contactless interactions.

We purposefully did not match the perceived intensities of mid-air ultrasound across the reference frequencies, nor the perceived intensities between ultrasound and mechanical vibrations, since this approach is impractical for haptic designers.
Our results informed our design of mid-air Tacton sets for the following studies, ensuring all the AM frequencies were above the JND threshold.

\section{Studies 2--3: Distinguishability of Mid-Air Tactons Varying on Parameters of Sinusoids}
\label{sec:Studies2-3}

We created two mid-air ultrasound Tacton sets corresponding to two mechanical Tacton sets that vary in parameters of sinusoids (Section~\ref{sec:TactonSetSelection}).
Then, we ran two user studies to derive the dissimilarity spaces for each mid-air Tacton set and compared the results to the dissimilarity spaces of the mechanical Tacton sets from the literature~\cite{park2011perceptual, lim2023can}.

\subsection{Mid-Air Ultrasound Tacton Design}
\label{sec:sinusoids}

We designed eight and twelve mid-air ultrasound Tactons for Studies 2 and 3 based on mathematical equations $x(t)$ that control the same parameters used for rendering mechanical Tactons~\cite{brewster2004tactons}.
In the equations, we defined $U(t)$ as the continuous ultrasound at 40\,kHz, $M(t)$ as an AM sinusoid of $sin(2\pi f_{AM} t)$ with AM frequency $f_{AM}$, and $E(t)$ as an envelope sinusoid of $sin(2\pi f_e t)$ with envelope frequency $f_e$.
Specifically, we used $f_{AM}$ as a corresponding parameter to the carrier frequency of the mechanical vibrations.
With this mapping, the AM frequency on the mid-air ultrasound device provides the same number of pulses on the palm as the AM frequency value (e.g., 10 pulses for a 10\,Hz frequency).
$A$ denotes the amplitude of ultrasound vibrations, where $A_{half}$ and $A_{full}$ correspond to 50\% and 100\% acoustic pressure on the mid-air ultrasound device, respectively.

In Study 2, we designed Tactons using the formula $x(t) = A_{full} \! \cdot \! U(t) M(t) E(t)$ (Figure~\ref{fig:Signal_UserStudy2}).
Tactons varied on eight envelope frequencies ($f_e$ = 0, 1, 2, 5, 10, 20, 40, or 80\,Hz) while AM frequency was constant at 150\,Hz.
$f_e$ = 0\,Hz represented a constant envelope (i.e., $E(t)$ = 1).
In Study 3, we designed Tactons using the formula $x(t) = A \! \cdot \! U(t) \{w_{\small AM_1} M_1 (t) + w_{AM_2} M_2 (t) \} E(t)$ (Figure~\ref{fig:Signal_UserStudy3}).
Tactons varied on two amplitudes ($A$ = $A_{half}$ or $A_{full}$), two envelope frequencies ($f_e$ = 0\,Hz or 4\,Hz), two AM frequencies ($f_{AM}$ = 70\,Hz or 210\,Hz), and three superposition ratios ($w_{AM_1}$:$w_{AM_2}$ = 1:0 (L), 0.5:0.5 (M), or 0:1 (H)).
In other words, L and H represented Tactons with 70\,Hz and 210\,Hz respectively while M was the superimposed Tacton with both 70\,Hz and 210\,Hz. 
The durations of all the Tactons in Studies 2 and 3 were 1 second.

\subsection{Participants}
We recruited 31 new participants for the two studies (n=15 per study), including eight females and 23 males, 20--34 years old.  
Each participant could only participate in one study.
One participant failed the attention test in Study 2, and their data was discarded.
The participants were all right-handed without any sensory impairments. 
On average, the participants took 29 and 49 minutes in Studies 2--3 and received a \$15 USD Amazon gift card.

\subsection{Experiment Procedure}
We followed the same procedure in both studies to collect pair-wise similarity ratings for mid-air ultrasound Tactons.
After obtaining informed consent, the participants completed a demographics pre-questionnaire, as in Study 1.
A 27" monitor displayed the GUI program in the studies. 
The training session displayed a set of buttons, each corresponding randomly to an ultrasound pattern (Figure~\ref{fig:rating_training}).
The participants experienced all the Tactons before the main session.

\begin{figure*}[t]

    \centering
  
    \begin{subfigure}[t]{1.0\textwidth}
    \includegraphics[width=\linewidth]{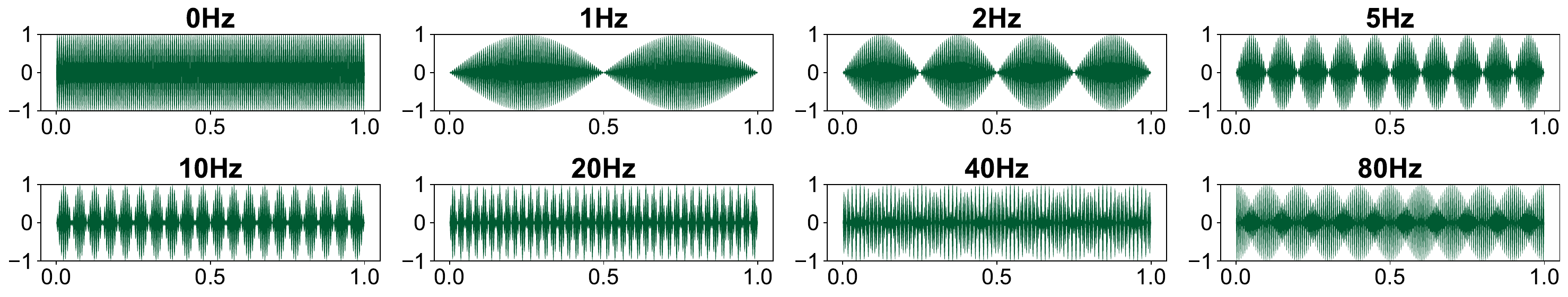}
    \end{subfigure}
    
    \begin{subfigure}[b]{1.0\textwidth}
    \caption{
    Eight mid-air ultrasound Tactons.
    We named the Tactons as \textbf{T-Envelope frequency\,(Hz)}. For example, \textbf{T-0} represents the Tacton with 0\,Hz envelope frequency.
    The x-axis represents time (seconds), and the y-axis corresponds to the acoustic pressure ranging between  0\%\ and $\pm$100\%\ on the STRATOS Explore device.
    }
    \label{fig:Signal_UserStudy2}
    \end{subfigure}

    \quad
    \quad
    \begin{subfigure}[t]{0.60\textwidth}
    \includegraphics[width=\linewidth]{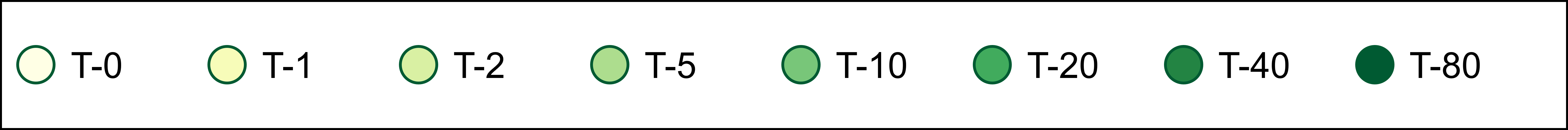}
    \end{subfigure}
    \hfill
    \begin{subfigure}[t]{0.30\textwidth}
    \includegraphics[width=\linewidth]{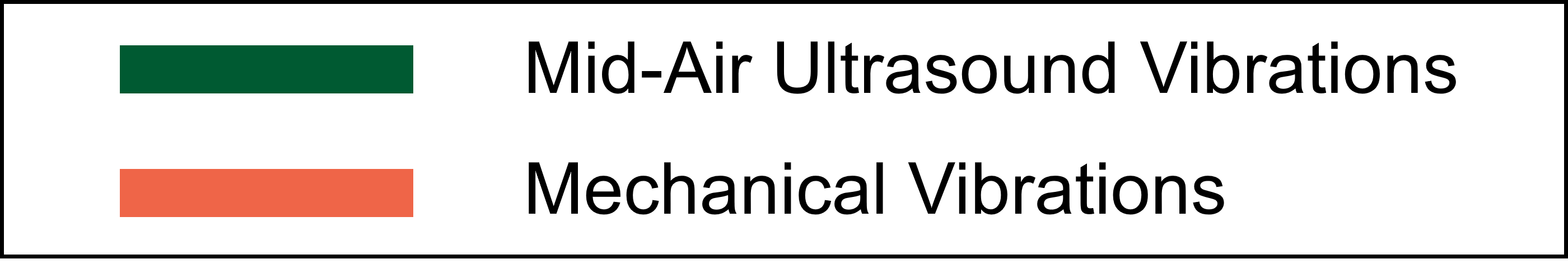}
    \end{subfigure}
    \quad

    \begin{subfigure}[b]{0.33\textwidth}
    \includegraphics[width=\linewidth]{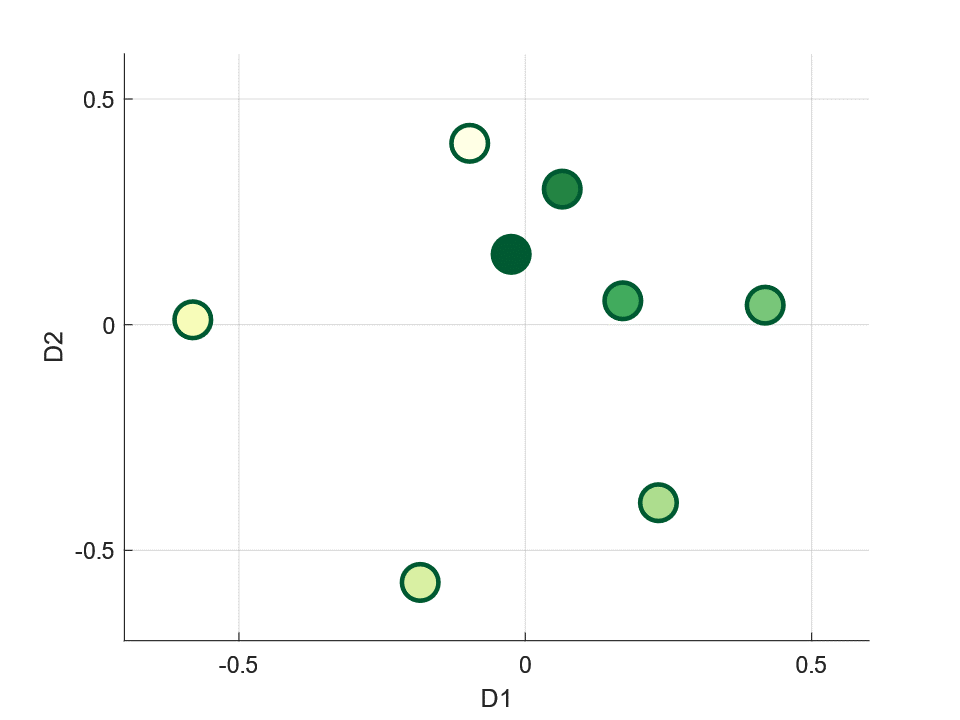}
    \end{subfigure}
    \hfill
    \begin{subfigure}[b]{0.33\textwidth}
    \includegraphics[width=\linewidth]{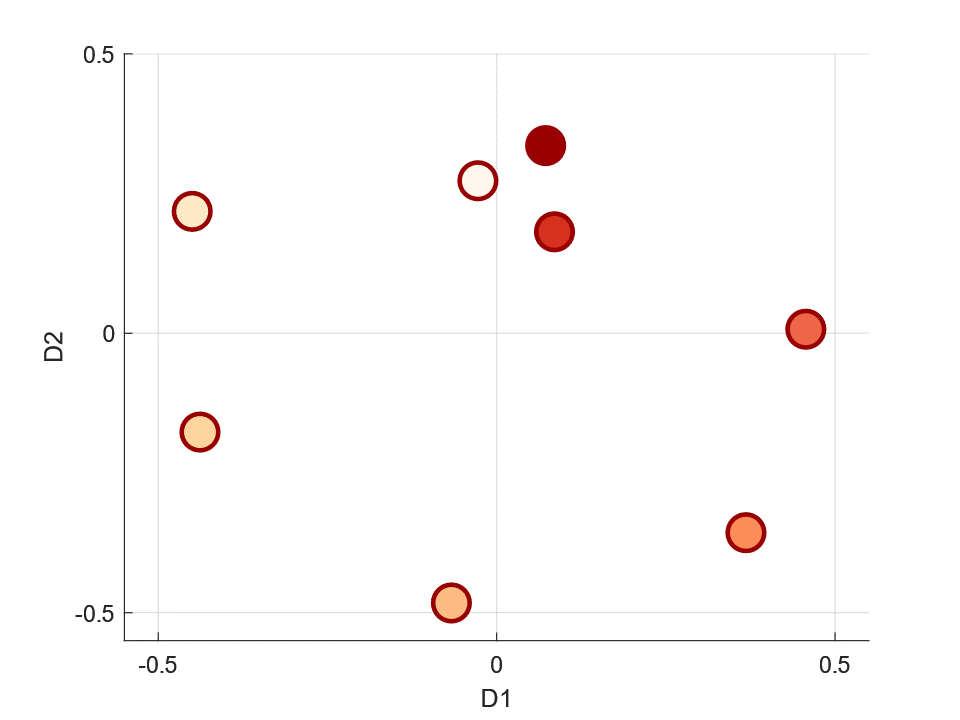}
    \end{subfigure}
    \hfill
    \begin{subfigure}[b]{0.33\textwidth}
    \includegraphics[width=\linewidth]{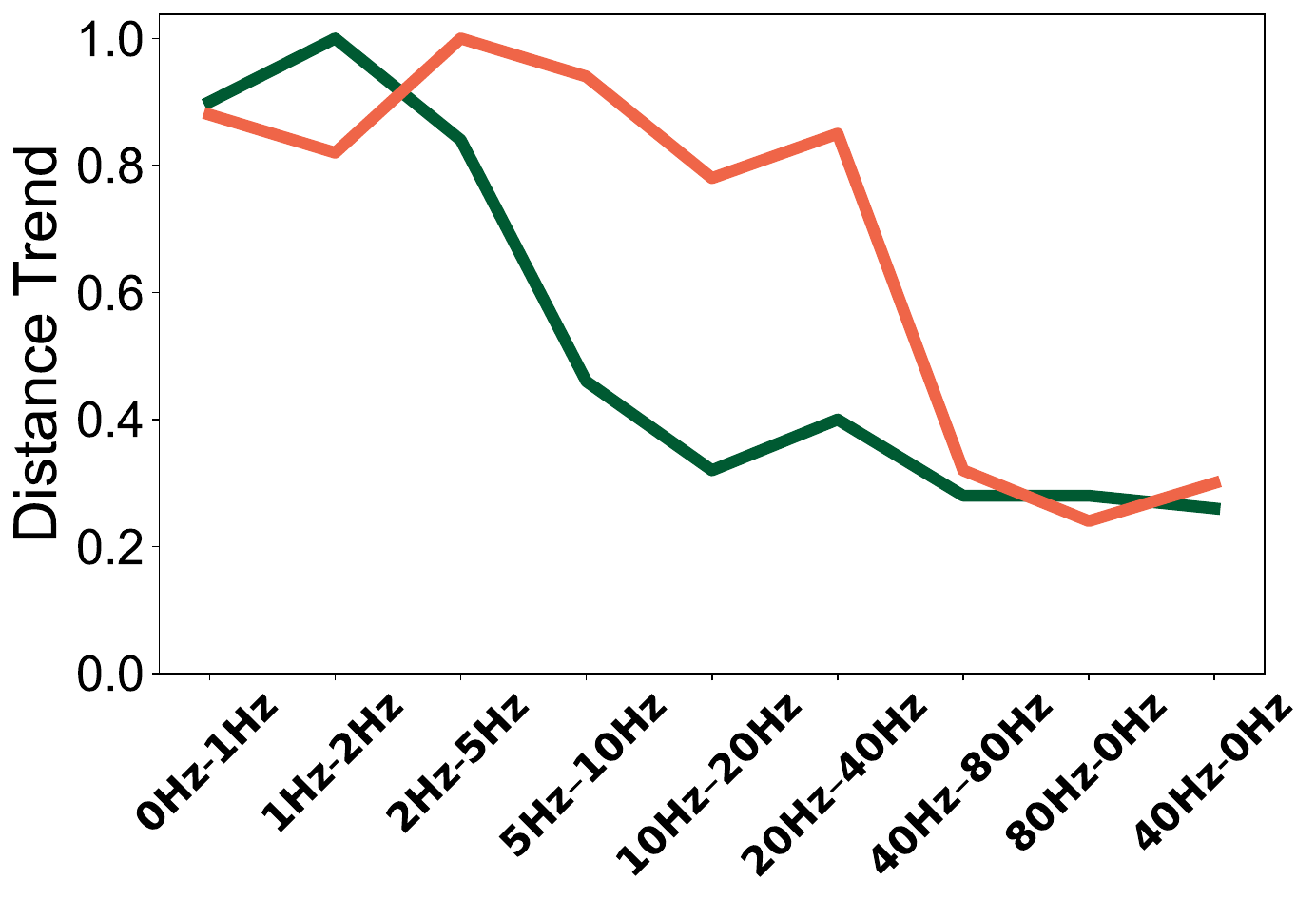}
    \end{subfigure}

    \begin{subfigure}[t]{0.66\textwidth}
    \caption{
    Perceptual spaces for mid-air ultrasound Tactons from our study (left) and mechanical Tactons from~\cite{park2011perceptual} (right) (Spearman's $\rho$ = 0.70).
    Kruskal's stress values were 0.03 and <0.01 for the perceptual spaces of ultrasound and mechanical Tactons, respectively, suggesting a good fit~\cite{kruskal1964multidimensional}.
    }
    \label{fig:PS_UserStudy2}
    \end{subfigure}
    \hfill
    \begin{subfigure}[t]{0.33\textwidth}
    \caption{Perceptual distance trends between adjacent envelope frequencies. The closer the value to 1, the more distinguishable the Tactons.}
    \label{fig:Trend_UserStudy2}
    \end{subfigure}

    \caption{ Plots of the ultrasound Tactons, perceptual spaces, and distance trends for Study 2.
    }
    \label{fig:Result_UserStudy2}
    \Description{Three plots labeled (a), (b), and (c) show signal plots of ultrasound Tactons, two perceptual spaces of ultrasound and mechanical Tactons, and distance trends in Study 2. Plot (a) shows the signal plots of eight ultrasound Tactons in the set, which vary in envelope frequency. Plot (b) shows a 2D perceptual space of the ultrasound Tactons on the left and a perceptual space of their corresponding mechanical Tactons on the right. The Tactons form a circular structure in both perceptual spaces as the envelope frequency increases, with minor differences. Plot (c) shows the perceptual distance trends between adjacent envelope frequencies in both ultrasound and mechanical Tactons. The perceptual distance decreases as the envelope frequency increases. For ultrasound Tactons, the distance shows a drastic decrease between 5Hz-10Hz, whereas for mechanical Tactons, the drastic decrease occurs between 20Hz-40Hz.}

\end{figure*}

In the main session, the participants rated the perceptual similarity for all possible pairs of the Tactons in a set (Figure~\ref{fig:rating_main}).
The main session presented each pair twice in random order and included an attention test with identical Tactons.
The participants rated the perceptual similarity of each pair of Tactons using a sliding bar ranging from 0 (totally different) to 100 (totally the same).
The participants could play the Tactons multiple times.
Finally, they completed a post-questionnaire as in Study 1.

\subsection{Analysis}
\label{sec:analysis}

We converted the similarity ratings to dissimilarity scores by subtracting them from 100 and calculated the average dissimilarity scores across 2 repetitions and 15 participants in each study. Then, we analyzed the similarity ratings for each study in two ways:
(1) We derived the dissimilarity space for each ultrasound Tacton set using a non-metric multidimensional scaling (nMDS) and analyzed the distances within the perceptual space, and 
(2) We compared the perceptual dissimilarities for the mid-air ultrasound Tactons with the dissimilarities for the corresponding mechanical Tacton set using Spearman rank correlation and visualization of the perceptual spaces.

The distances in two perceptual spaces are not directly comparable because the user similarity ratings are relative to the variations in a Tacton set.
Thus, when reporting the comparisons between the mid-air ultrasound and mechanical Tactons, we first selected the target parameters that we want to compare in perceptual spaces.
Then, we normalized the perceptual distances for the selected parameters by dividing them with the maximum distance.
We compared the normalized distances to discuss the trends between the Tacton sets in the two modalities.

For both analyses, we followed the established procedure and metrics (e.g., nMDS, Spearman correlation) from the literature on mechanical Tactons \cite{park2011perceptual,ternes2008designing,abou2022vibrotactile,kwon2023can}. 
We visualized the perceptual space of ultrasound Tacton sets using the same number of dimensions proposed for the corresponding mechanical Tacton sets in the literature.
We also reported Kruskal's stress values to indicate the goodness of fit of the visualizations for the ultrasound Tactons~\cite{kruskal1964multidimensional}.

\subsection{Study 2 Results}
\label{sec:Study2_Results}

\textbf{Distinguishability of mid-air ultrasound Tactons:} 
The eight Tactons, varied by envelope frequency, formed an anticlockwise circle in the perceptual space starting from \textbf{T-0} to \textbf{T-80} (Figure~\ref{fig:PS_UserStudy2}). 
The perceptual distances between adjacent frequencies showed a large drop after 5\,Hz (Figure~\ref{fig:Trend_UserStudy2}).
When the envelope frequencies reached 40\,Hz and 80\,Hz, the Tactons (\textbf{T-40} and \textbf{T-80}) became difficult to distinguish from the Tacton with a constant envelope (\textbf{T-0}).
Overall, the perceptual distance between lower envelope frequency pairs ($\leq$ 5\,Hz) was larger than the distance between higher envelope frequency pairs ($\geq$ 10\,Hz) in the Tacton set.

\begin{figure*}[t]

    \centering
  
    \begin{subfigure}[t]{1.0\textwidth}
    \includegraphics[width=\linewidth]{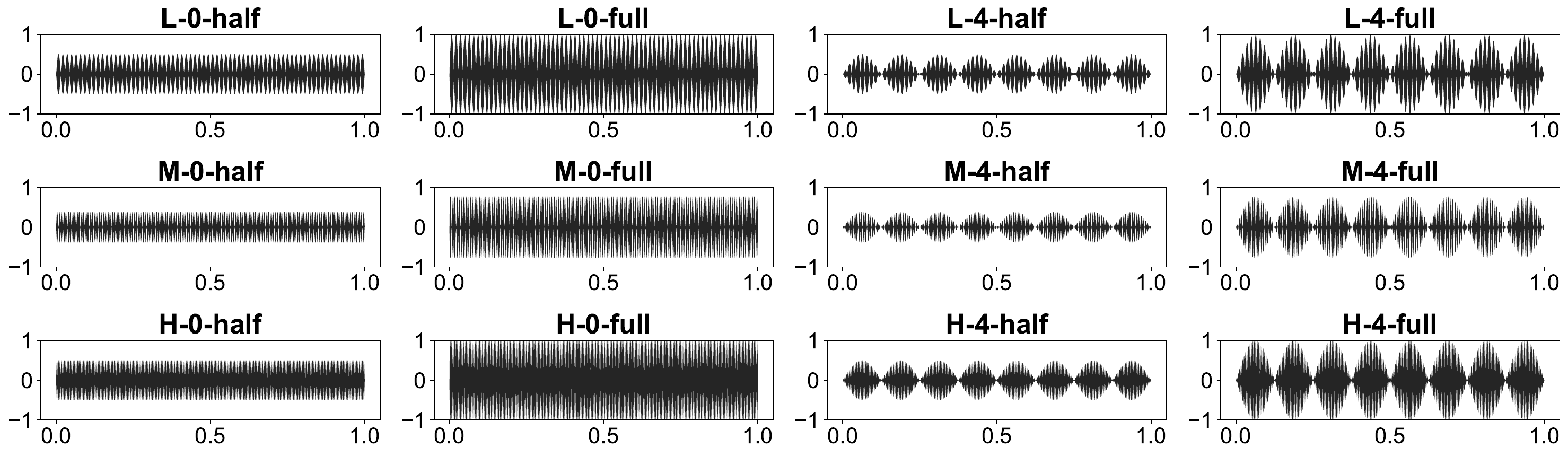}
    \end{subfigure}
    
    \begin{subfigure}[b]{1.0\textwidth}
    \caption{
    Twelve mid-air ultrasound Tactons.
    We named Tactons as \textbf{Superposition ratio - Envelope frequency - Amplitude}.
    }
    \label{fig:Signal_UserStudy3}
    \end{subfigure}

    \quad
    \quad
    \begin{subfigure}[c]{0.6\textwidth}
    \includegraphics[width=\linewidth]{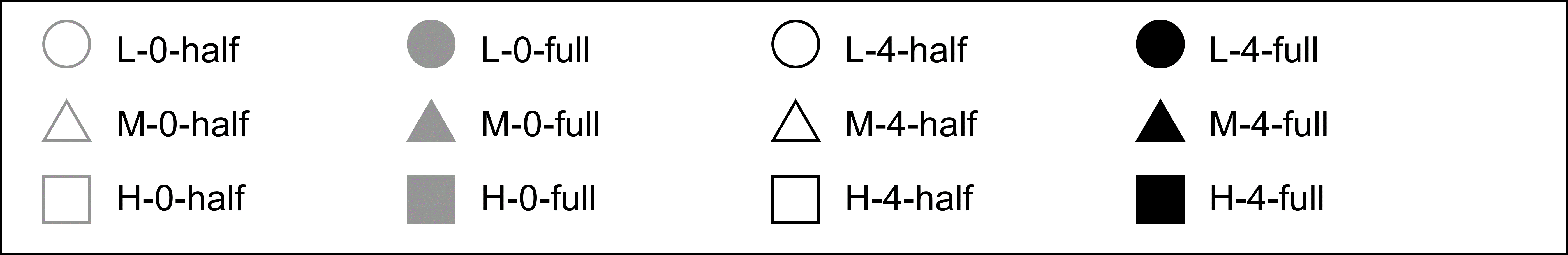}
    \end{subfigure}
    \hfill
    \begin{subfigure}[c]{0.3\textwidth}
    \includegraphics[width=\linewidth]{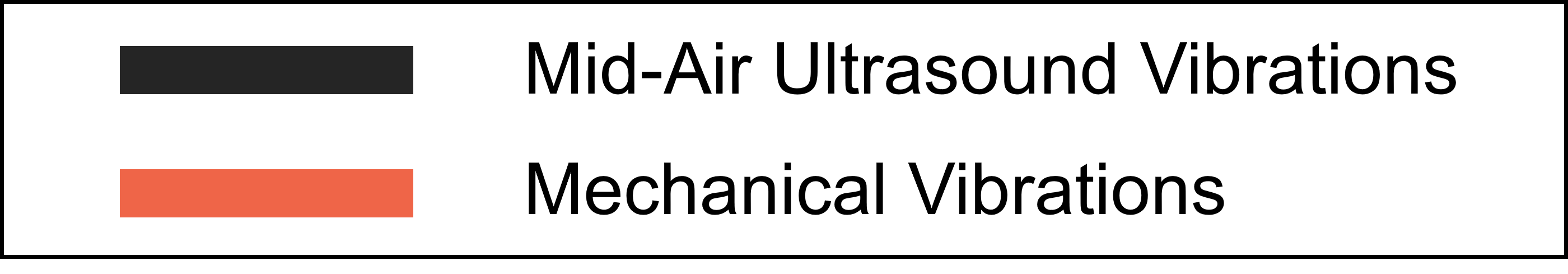}
    \end{subfigure}
    \quad

    \begin{subfigure}[c]{0.33\textwidth}
    \includegraphics[width=\linewidth]{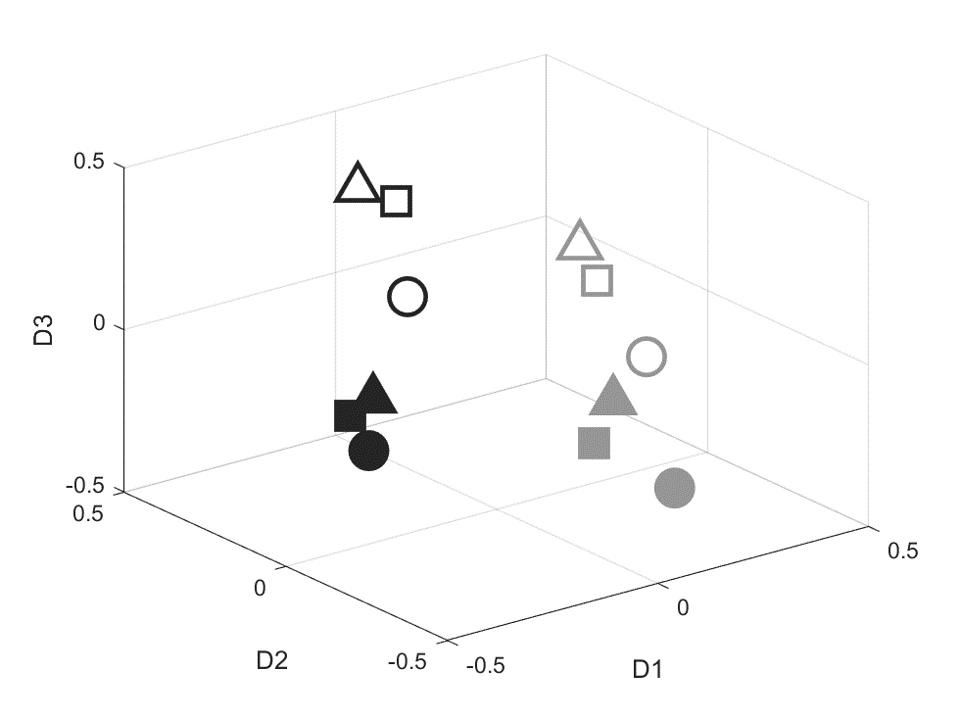}
    \end{subfigure}
    \hfill
    \begin{subfigure}[c]{0.33\textwidth}
    \includegraphics[width=\linewidth]{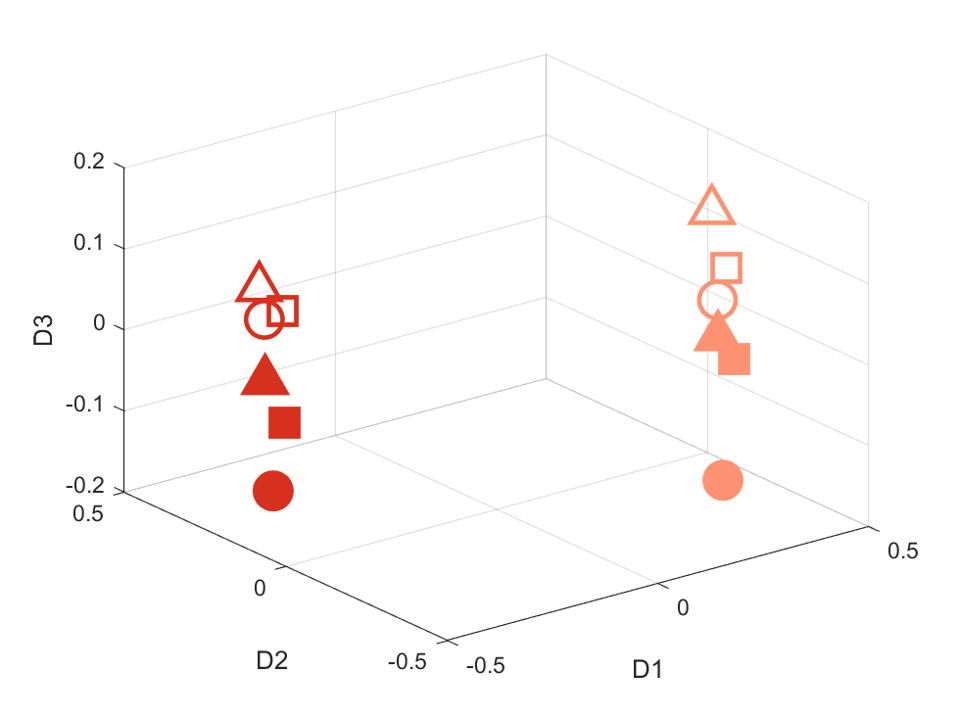}
    \end{subfigure}
    \hfill
    \hfill
    \begin{subfigure}[c]{0.33\textwidth}
    \includegraphics[width=\linewidth]{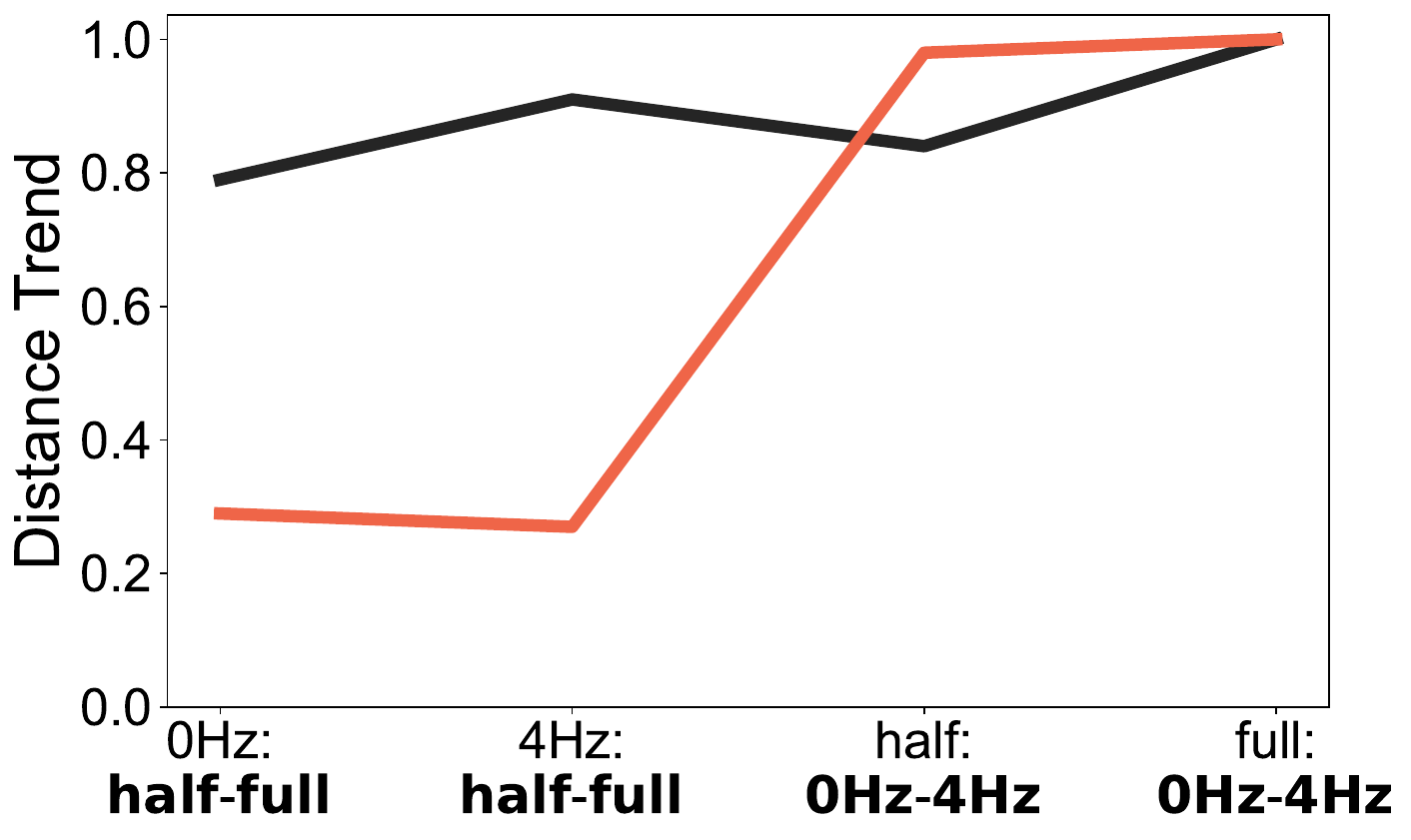}
    \end{subfigure}

    \begin{subfigure}[t]{0.66\textwidth}
    \caption{Perceptual spaces for mid-air ultrasound Tactons from our study (left) and mechanical Tactons from~\cite{lim2023can} (right) (Spearman's $\rho$ = 0.61). The Kruskal's stress values were 0.03 for both perceptual spaces, suggesting a good fit.}
    \label{fig:PS_UserStudy3}
    \end{subfigure}
    \hfill
    \begin{subfigure}[t]{0.33\textwidth}
    \caption{Perceptual distance trends between adjacent clusters formed by mixing amplitudes and envelope frequencies.}
    \label{fig:Trend_UserStudy3_Clusters}
    \end{subfigure}

    \caption{
    Plots of the ultrasound Tactons, perceptual spaces, and distance trends for Study 3.
    }
    \label{fig:Result_UserStudy3}
    \Description{The three plots labeled (a), (b), and (c) show signal plots of ultrasound Tactons, two perceptual spaces of ultrasound and mechanical Tactons, and distance trends in Study 3. Plot (a) shows the signal plots of twelve ultrasound Tactons in the set, varying in envelope frequency, amplitude, amplitude-modulated frequency, and superposition ratio. Plot (b) shows a 3D perceptual space for the ultrasound Tactons on the left and a corresponding perceptual space for the mechanical Tactons on the right. In both spaces, the Tactons form four clusters determined by amplitude and envelope frequency. They also form similar local distributions in each cluster with minor differences. Plot (c) shows the perceptual distance trends for adjacent clusters by envelope frequency and amplitude in ultrasound and mechanical Tactons. For ultrasound Tactons, the perceptual distances between adjacent clusters, whether formed by amplitude or envelope frequency, are similar. In contrast, for mechanical Tactons, the distances between clusters formed by amplitude are smaller than those determined by envelope frequency.}

\end{figure*}

\textbf{Comparison with mechanical vibrations:} 
The dissimilarity values between mid-air ultrasound and mechanical Tactons were significantly correlated with Spearman's $\rho = 0.70$ ($p<0.001$). A Spearman $\rho$ above 0.6 is considered a strong correlation in the literature~\cite{chan2003biostatistics}. Also, the visualization of perceptual spaces show similar configuration for ultrasound and mechanical Tactons (Figure~\ref{fig:PS_UserStudy2}).
We observed similar trends between lower envelope frequency pairs ($\leq$ 5\,Hz) in Figure~\ref{fig:Trend_UserStudy2}.
Moreover, in both mid-air ultrasound and mechanical vibrations, \textbf{T-40} and \textbf{T-80} were difficult to distinguish from \textbf{T-0}.
However, the drastic drop in perceptual distance occurred at lower frequencies in mid-air ultrasound Tactons (5\,Hz) than in mechanical Tactons (40\,Hz).

\textbf{Brief discussion:} 
The results demonstrate the human ability to perceive the number of pulses in the low envelope frequency ranges ($f_e \leq$ 5\,Hz) for both mid-air ultrasound and mechanical vibrations~\cite{park2011perceptual, brown2005first}.
Therefore, haptic designers can create ultrasound Tactons with countable pulses using the low envelope frequencies ($f_e \leq$ 5\,Hz).
Although the frequency spectra of signals with a high envelope frequency ($\geq$ 40\,Hz) differed from that of a constant-envelope signal (0\,Hz), users rated the Tactons with high $f_e$ (i.e., \textbf{T-40} and \textbf{T-80}) as the most similar to \textbf{T-0}. 
This result suggests that the temporal characteristics (i.e., temporal envelope or pulse count) are perceptually more salient than the spectral characteristics for both the ultrasound and mechanical Tactons in this set~\cite{park2011perceptual}.

\subsection{Study 3 Results}

\textbf{Distinguishability of mid-air ultrasound Tactons:} 
In the derived perceptual space (Figure~\ref{fig:PS_UserStudy3}-left), the Tactons were separated by envelope frequency (0\,Hz / 4\,Hz), followed by amplitude (half / full). The superposition ratio (L / M / H) created the lowest perceptual distance. The envelope frequency was aligned with D1 and D2, while amplitude was aligned with D3, segmenting the perceptual space into four clusters. 
In each cluster, the AM frequency (70\,Hz / 210\,Hz) and superposition ratio (L / M / H) formed a local distribution of M-H-L from top to bottom.
These results imply that envelope frequency and amplitude were the primary contributors to distinguishability for this set, while the AM frequency and superposition ratio acted as secondary parameters.
In all the local distributions within each cluster, users rated that the 70\,Hz Tacton (L) was more distinguishable from the superimposed (M) and 210\,Hz (H) Tactons while M and H were perceptually similar.

\textbf{Comparison with mechanical vibrations:} 
The perceptual dissimilarities between mid-air ultrasound Tactons and mechanical Tactons had a strong correlation (Spearman's $\rho = 0.61$, $p<0.001$).
The visualizations of both perceptual spaces showed four clusters divided by amplitude and envelope frequency.
In all four clusters of both perceptual spaces, frequency and superposition ratio formed a similar local distribution of M-H-L, and the superimposed Tactons (M) were consistently positioned outside the distance between the high-frequency (H) and low-frequency Tactons (L).

However, the perceptual distances between adjacent clusters showed a different trend in the two technologies.
For mid-air ultrasound Tactons, the effect of the amplitude on Tacton distances was similar to the effect of envelope frequency (0.85 and 0.92 in Figure~\ref{fig:Trend_UserStudy3_Clusters}).
However, for mechanical Tactons, Lim and Park reported that Tactons with different envelope frequencies had higher distance (0.99) than Tactons with different amplitudes (0.28).

To interpret the larger perceptual differences between half and full amplitudes in ultrasound compared to mechanical vibrations, we followed established analysis methods that calculate sensation levels within a single modality such as mid-air ultrasound~\cite{verrillo1992perception, ryu2010psychophysical}.
Here, the sensation level is defined by $dB = 20 log(A/AL)$, where $A$ is the physical amplitude and $AL$ is the corresponding absolute detection threshold.
Based on the amplitude detection threshold of mid-air ultrasound vibrations on the palm~\cite{howard2019investigating}, we derived the approximate sensation levels to be about 4.5\,$dB$ for half (562.5\,Pa) and 10.5\,$dB$ for full (1125\,Pa) amplitudes for our Tactons.
Thus, the sensation level of half in mid-air ultrasound vibrations was close to the human amplitude detection threshold and about 57\% of the sensation level for the full amplitude. 
In contrast, for mechanical vibrations~\cite{lim2023can}, the sensation levels for half (0.8\,G) and full (1.6\,G) amplitudes at 70\,Hz were around 40.2\,$dB$ and 46.2\,$dB$, respectively~\cite{ryu2010psychophysical}.
At 210\,Hz mechanical vibrations, these levels were approximately 42.1\,$dB$ and 48.1\,$dB$.
The sensation levels of both half and full amplitudes were above the detection threshold for mechanical vibrations~\cite{ryu2010psychophysical}, and their drop ratios were both 13\% at 70\,Hz and 210\,Hz.

\textbf{Brief discussion:} 
The envelope frequencies (0\,Hz and 4\,Hz) and two amplitude levels (half and full) contributed more to perceptual distances than the AM frequency and superposition ratio in this Tacton set.
The envelope frequency and amplitude were effective in both mid-air ultrasound and mechanical Tactons, while their relative impacts on perceptual distances varied.
Our results suggest the same ratio in amplitude (half and full) for mid-air ultrasound can lead to a larger drop in the sensation level and produce greater perceptual distances between half and full in ultrasound than in mechanical vibrations.

The effect of the superposition ratio on the perceptual distances of ultrasound Tactons was consistent across all four clusters in both ultrasound and mechanical Tactons. 
In both cases, the superimposed Tacton (M) fell outside the two pure sinusoids (L and H). Although the perceptual distances were small, this result suggests the superposition ratio may create qualitatively different sensations.

\section{Study 4: Distinguishability of Mid-Air Tactons Varying on Rhythms}
\label{sec:Study4}

In Study 4, we examined the perceptual distinguishability for mid-air ultrasound Tactons that were inspired by musical structures and varied in rhythm and amplitude (Section~\ref{sec:TactonSetSelection}), based on a mechanical Tacton set from the literature~\cite{abou2022vibrotactile}.
The study setup, procedure, and analysis were the same as in Studies 2--3.

\begin{figure*}[t]

    \centering
  
    \begin{subfigure}[t]{1.0\textwidth}
    \includegraphics[width=\linewidth]{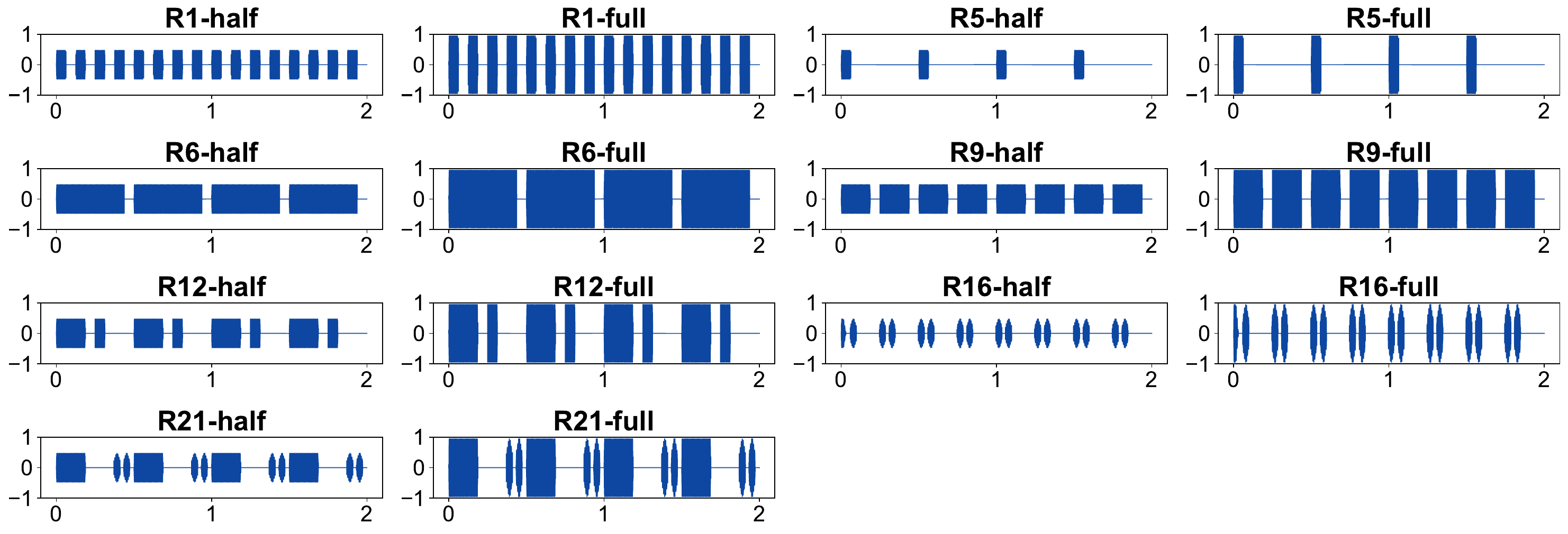}
    \end{subfigure}
    
    \begin{subfigure}[b]{1.0\textwidth}
    \caption{
    Fourteen mid-air ultrasound Tactons.
    We named Tactons as \textbf{Rhythm - Amplitude}.
    }
    \label{fig:Signal_UserStudy4}
    \end{subfigure}

    \begin{subfigure}[t]{0.7\textwidth}
    \includegraphics[width=\linewidth]{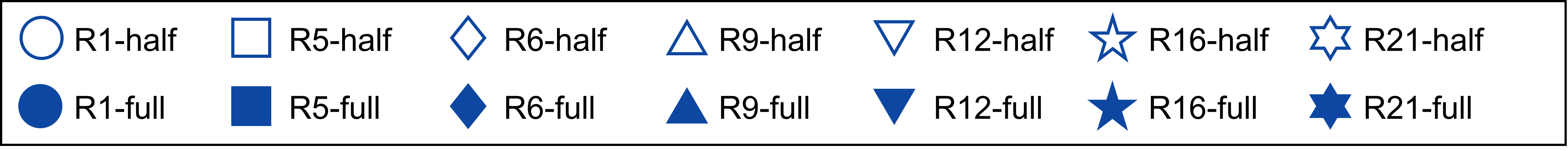}
    \end{subfigure}

    \begin{subfigure}[t]{0.36\textwidth}
    \includegraphics[width=\linewidth]{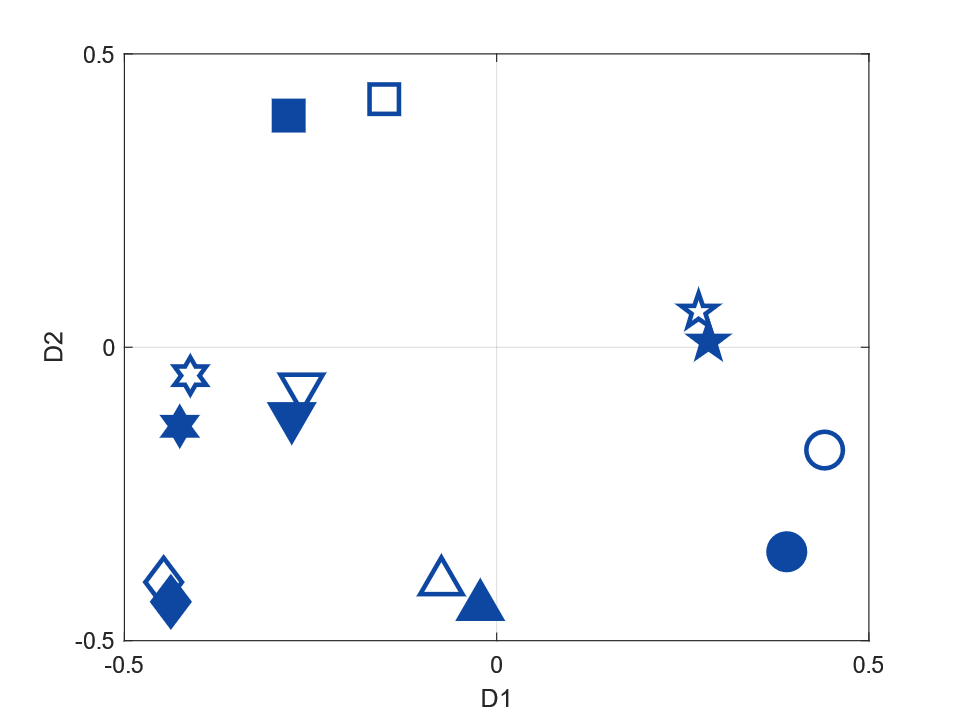}
    \end{subfigure}
    \quad
    \begin{subfigure}[t]{0.36\textwidth}
    \includegraphics[width=\linewidth]{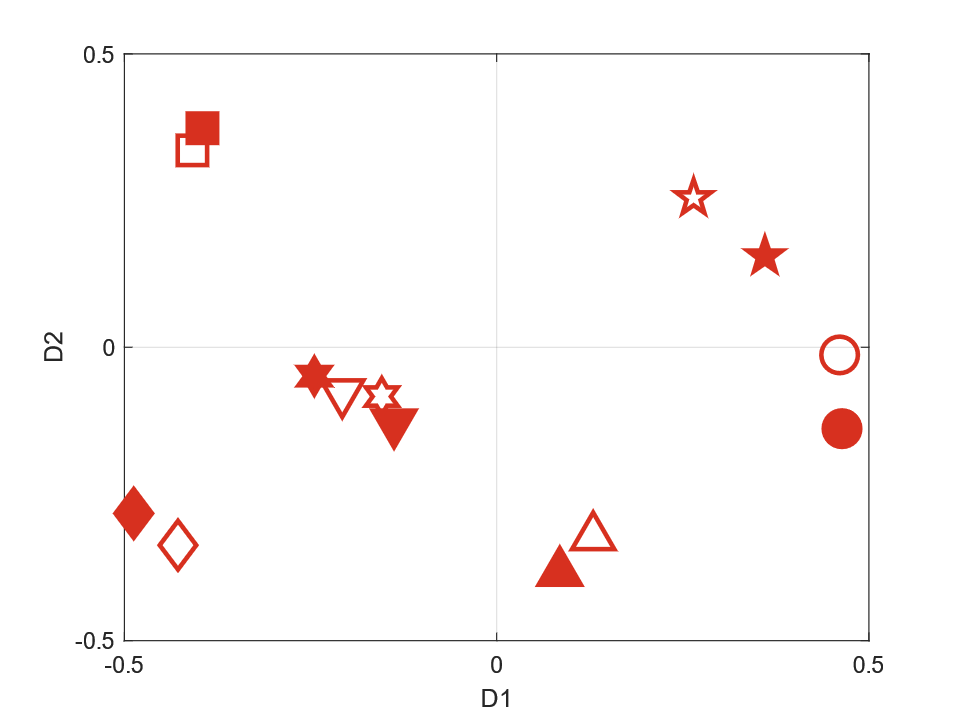}
    \end{subfigure}

    \begin{subfigure}[t]{1.0\textwidth}
    \caption{Perceptual spaces for mid-air ultrasound Tactons from our study (left) and mechanical Tactons from~\cite{abou2022vibrotactile} (right). The dissimilarity values for the ultrasound and mechanical Tactons had a very strong correlation (Spearman's $\rho$ = 0.89). The Kruskal's stress values were 0.10 for both perceptual spaces, suggesting a fair fit~\cite{kruskal1964multidimensional}.}
    \label{fig:PS_UserStudy4}
    \end{subfigure}

    \caption{Plots of the mid-air ultrasound Tactons and the perceptual spaces for Study 4.
    }
    \label{fig:Result_UserStudy4}
    \Description{The two plots labeled (a) and (b) show signal plots of ultrasound Tactons and two perceptual spaces of ultrasound and mechanical Tactons in Study 4. Plot (a) shows the signal plots of fourteen ultrasound Tactons in the set, varying in rhythm and amplitude. Plot (b) shows a 2D perceptual space for the ultrasound Tactons on the left and a corresponding perceptual space for the mechanical Tactons on the right. The Tactons form a circular structure in both perceptual spaces by rhythm.}
    
\end{figure*}

\subsection{Mid-Air Ultrasound Tacton Design}
\label{sec:rhythm}
We designed 14 mid-air ultrasound Tactons using the formula $x(t) = A \! \cdot \! U(t) M(t) R(t)$ (Figure~\ref{fig:Signal_UserStudy4}).
$A$, $U(t)$, and $M(t)$ were the same as in Section \ref{sec:sinusoids}, and we defined $R(t)$ as a rhythmic envelope from a past study~\cite{abou2022vibrotactile}.
We used seven rhythms ($R(t)$: R1, R5, R6, R9, R12, R16, or R21) and two amplitudes ($A$ = $A_{half}$ or $A_{full}$).
The AM frequency was constant at 150\,Hz.
We applied a 40\,ms moving average to the notes of $R(t)$ whose lengths were lower than 40\,ms.
This step was based on a pilot study with five people where we found pulses less than 40\,ms were hard to perceive.
All the Tactons were 2 seconds.

\subsection{Participants}
We recruited 15 new participants for Study 4 (three females and 12 males, 21--26 years old), including 4 left-handed and 11 right-handed users with no sensory impairments in their hands. 
On average, the participants took 76 minutes and received a \$22 USD Amazon gift card.

\subsection{Study 4 Results}
\textbf{Distinguishability of mid-air ultrasound Tactons:} 
The rhythm formed a circular distribution in the derived perceptual space, suggesting its potential in designing distinct Tactons.
Furthermore, the rhythm was more salient than the amplitude in perceiving the similarity of mid-air ultrasound Tactons. 
Specifically, the ultrasound Tactons were structured according to their number of notes (i.e., pulses) (D1) and the total length of notes (D2).
Along D1, the number of notes increased from Tactons on the left (negative D1) to Tactons on the right (Figure \ref{fig:PS_UserStudy4}-left).
The position of Tactons along D1 showed a strong correlation with the number of notes (Pearson's $r = 0.75$, $p<0.01$).
Along D2, the total length of notes decreased from the Tactons on the bottom (negative D2) to the Tactons on the top.
The position of Tactons along D2 had a very strong correlation with the total length of notes ($r = -0.94$, $p<0.01$)~\cite{chan2003biostatistics}.

\textbf{R6} and \textbf{R16} were the most different perceptually.
\textbf{R6} had 4 notes with a total note length of 1.75 seconds, while \textbf{R16} had 16 notes with a total note length of approximately 0.64 seconds.
\textbf{R12} and \textbf{R21} were perceived as the most similar.
\textbf{R12} and \textbf{R21} had similar total note lengths of 1 and 1.07 seconds, respectively, with slightly different numbers of notes.
Moreover, both \textbf{R12} and \textbf{R21} had a combination of short and long notes, providing a sense of unevenness to users~\cite{ternes2008designing}. 
Compared to the rhythm, the perceptual distances from the amplitude parameter were small in this Tacton set.

\textbf{Comparison with mechanical vibrations:}
The dissimilarity values for mid-air ultrasound Tactons showed a very strong correlation (Spearman's $\rho$ = 0.89, $p<0.001$)~\cite{chan2003biostatistics} with the dissimilarity values for the mechanical Tactons in the literature~\cite{abou2022vibrotactile}.
In both Tacton sets, rhythm dominates over amplitude.
Similar to the ultrasound technology, the mechanical Tactons were structured according to their number of notes (D1, $r = 0.87$, $p<0.01$) and the total length of notes (D2, $r = -0.96$, $p<0.01$)~\cite{chan2003biostatistics}.
Participants consistently reported that \textbf{R6} and \textbf{R16} were the most different rhythms, while \textbf{R12} and \textbf{R21} were the most similar rhythms across both Tacton sets.

\textbf{Brief discussion:}
Our results suggest that rhythm has a strong impact on the distinguishability of both mid-air ultrasound and mechanical Tactons.
In particular, the number and total length of notes served as the primary perceptual dimensions for rhythmic Tactons in both modalities.
Furthermore, in both cases, the note length (long vs. short) and evenness (even vs. uneven) impacted the perceptual spaces of the Tactons, making rhythm a robust design parameter for creating Tactons across contact-based and contactless interactions.

\section{Study 5: Distinguishability of Mid-Air Tactons Varying on Complex Waveforms}
\label{sec:Study5}

Study 5 examined the distinguishability of 14 mid-air ultrasound Tactons that vary on their target metaphors (metaphor-based design)  based on mechanical Tactons from an existing vibration library~\cite{seifi2015vibviz}.
The study setup, procedure, and analysis were the same as in Studies 2--4.

\begin{figure*}[t]

    \centering
  
    \begin{subfigure}[t]{1.0\textwidth}
    \includegraphics[width=\linewidth]{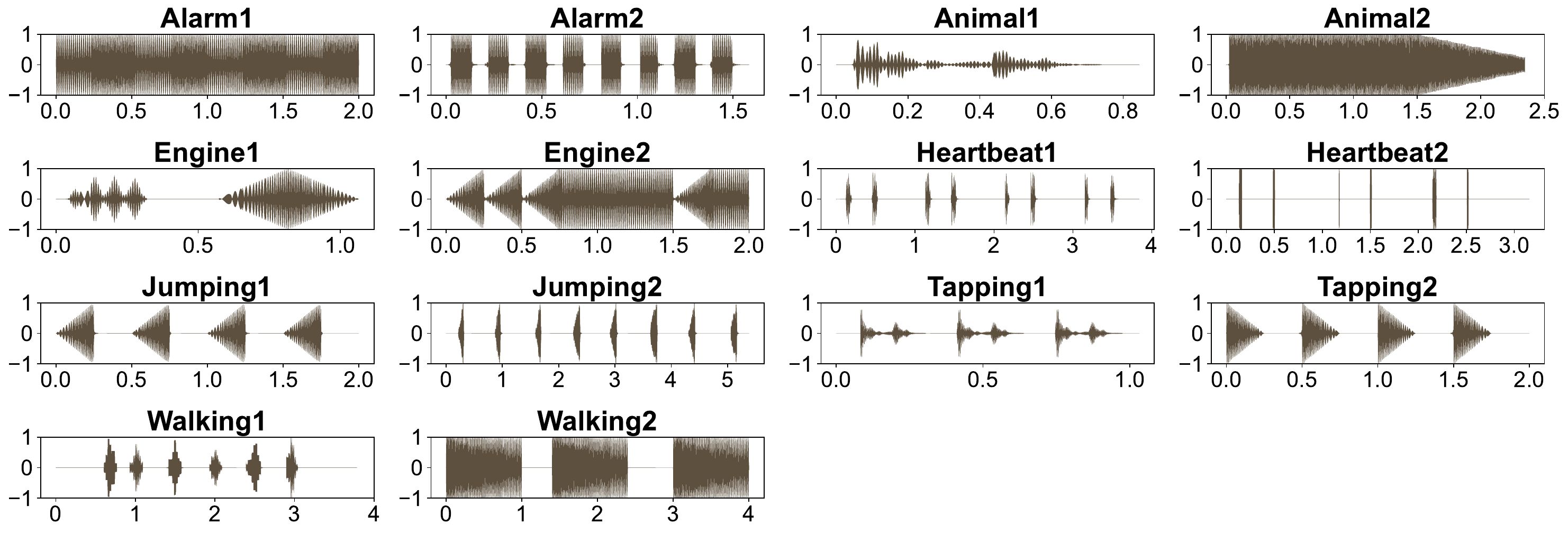}
    \end{subfigure}
    
    \begin{subfigure}[b]{1.0\textwidth}
    \caption{Fourteen mid-air ultrasound Tactons.
    The Tacton names are the same as a past study~\cite{kwon2023can}.
    }
    \label{fig:Signal_UserStudy5}
    \end{subfigure}

    \begin{subfigure}[t]{0.7\textwidth}
    \includegraphics[width=\linewidth]{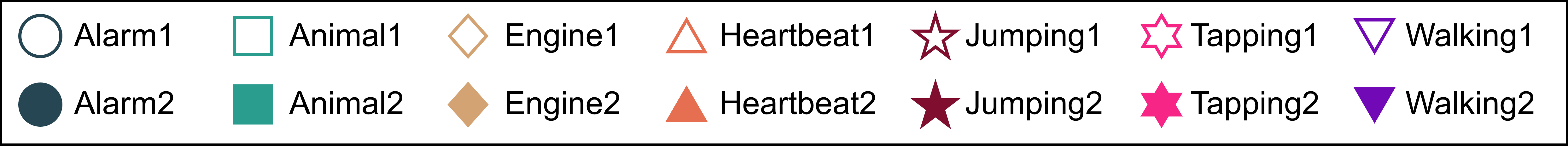}
    \end{subfigure}

    \begin{subfigure}[t]{0.36\textwidth}
    \includegraphics[width=\linewidth]{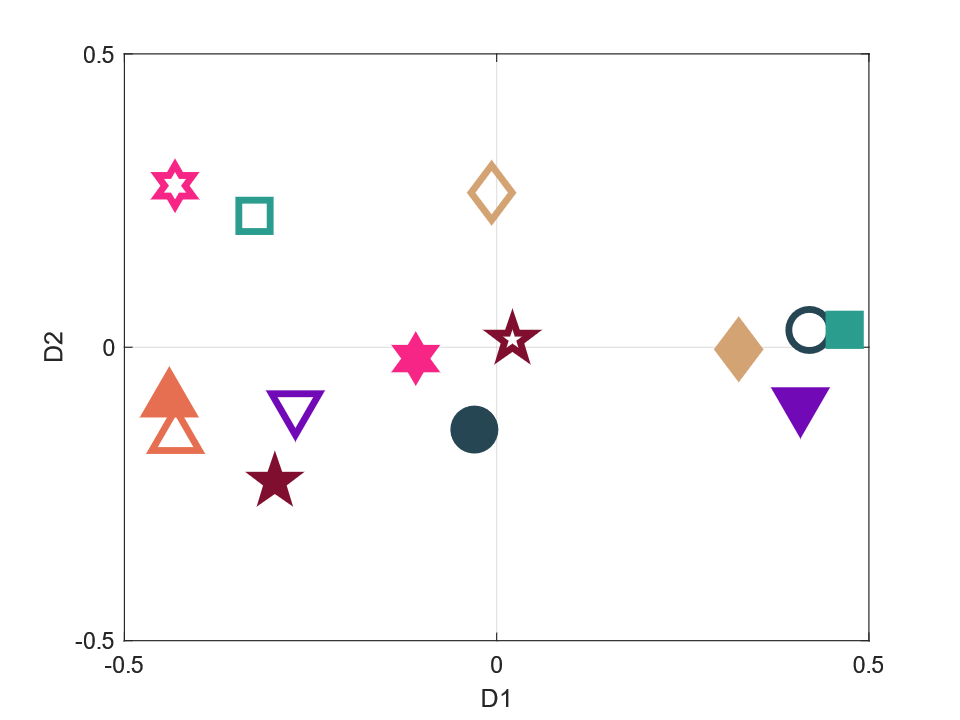}
    \end{subfigure}
    \quad
    \begin{subfigure}[t]{0.36\textwidth}
    \includegraphics[width=\linewidth]{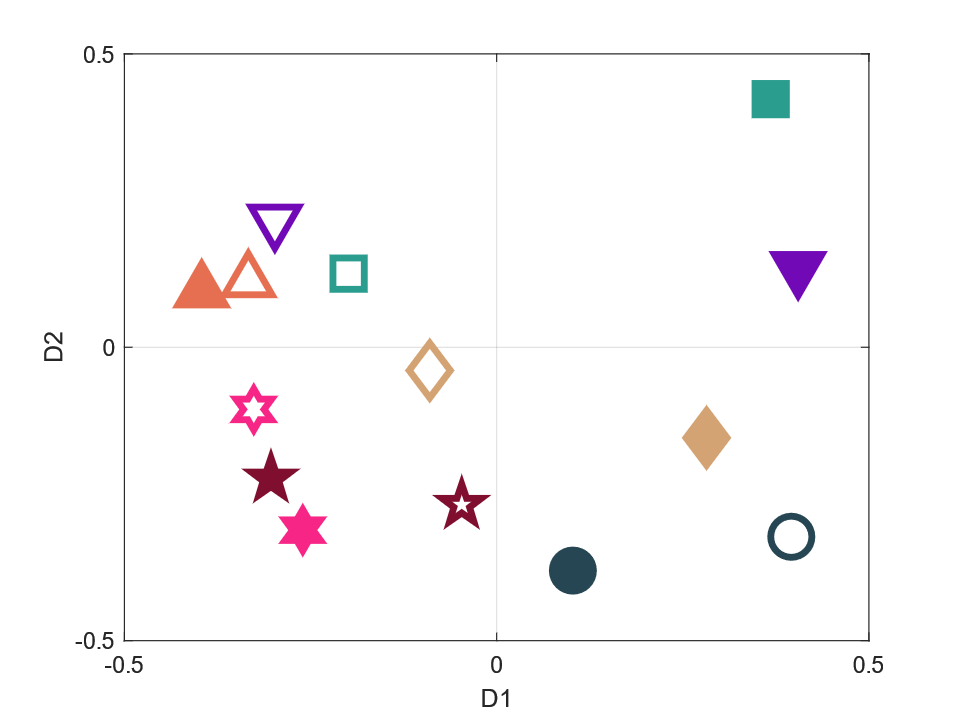}
    \end{subfigure}

    \begin{subfigure}[t]{0.36\textwidth}
    \includegraphics[width=\linewidth]{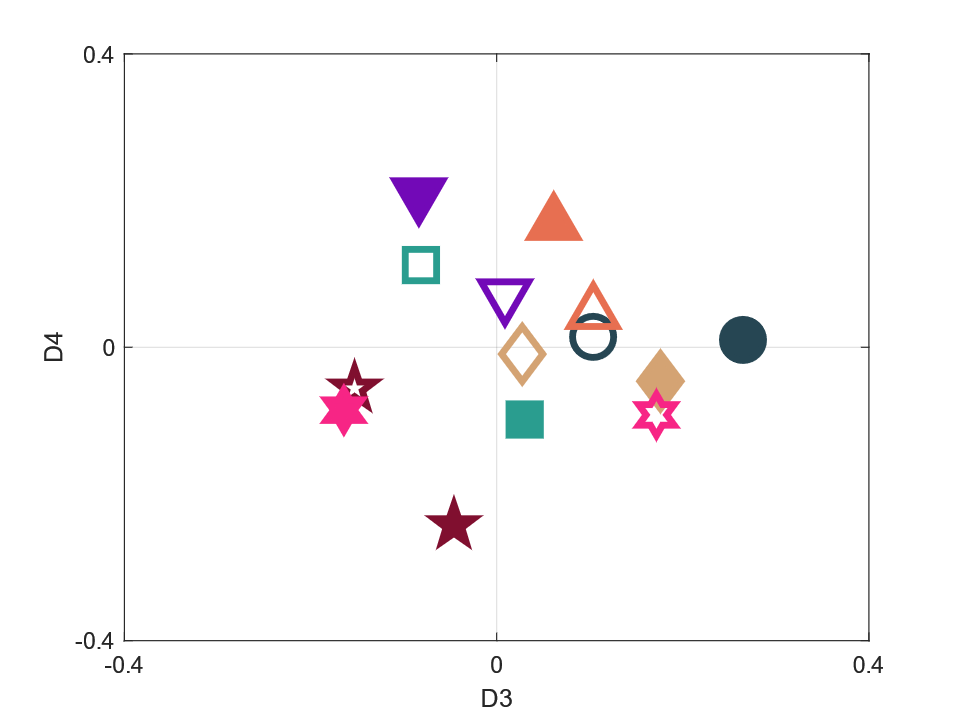}
    \end{subfigure}
    \quad
    \begin{subfigure}[t]{0.36\textwidth}
    \includegraphics[width=\linewidth]{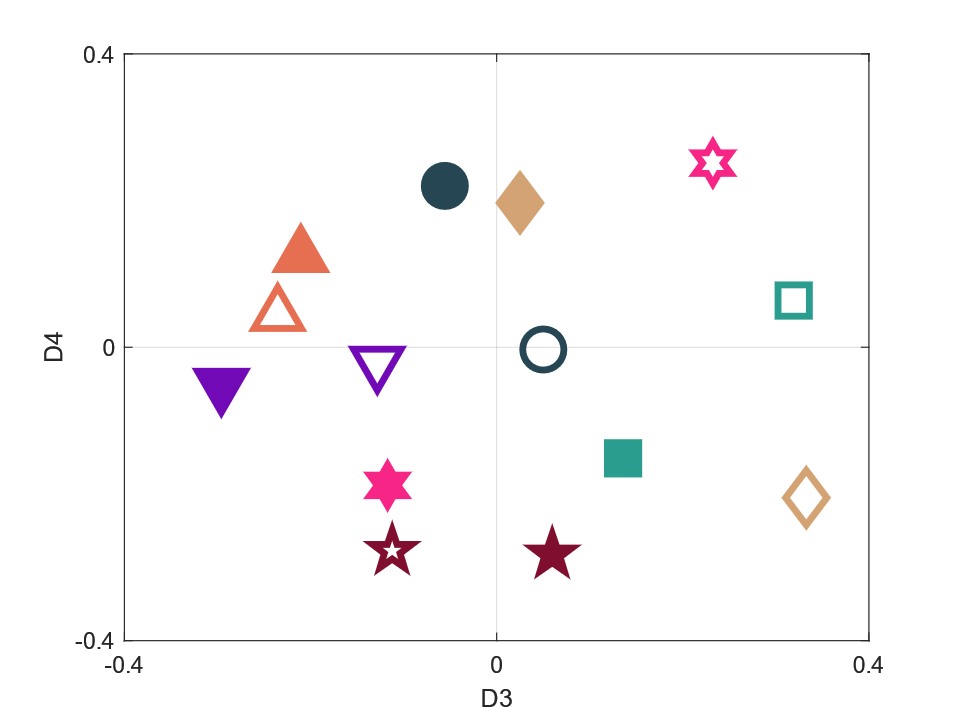}
    \end{subfigure}

    \begin{subfigure}[t]{1.0\textwidth}
    \caption{Perceptual spaces for mid-air ultrasound Tactons from our study (left) and mechanical Tactons from~\cite{kwon2023can} (right). The user ratings showed a strong correlation between the perceived Tacton dissimilarities in the two technologies (Spearman's $\rho$ = 0.78). The Kruskal's stress values suggested four dimensions provided a good fit for the perceptual spaces of ultrasound (0.05) and mechanical Tactons (0.02), respectively.}
    \label{fig:PS_UserStudy5}
    \end{subfigure}

    \caption{ Plots of the mid-air ultrasound Tactons and the perceptual spaces for Study 5.
    }
    \Description{The two plots labeled (a) and (b) show signal plots of ultrasound Tactons and four perceptual spaces of ultrasound and mechanical Tactons in Study 5. Plot (a) shows the signal plots of fourteen ultrasound Tactons in the set, varying in metaphor. Plot (b) shows 4D perceptual spaces of ultrasound and mechanical Tactons, but they are visualized as two 2D perceptual spaces of D1-D2 and D3-D4, respectively. Perceptual spaces for the ultrasound Tactons are on the left and corresponding perceptual spaces for the mechanical Tactons on the right. In the perceptual space of ultrasound Tactons, Tactons are structured according to their rhythmic structure (D1) and the total duration of the Tacton (D2). In the perceptual space of mechanical Tactons, Tactons are structured according to their rhythmic structure (D1), the change in frequency spectrum (D2), and the total duration of the Tacton (D3).}
    
\end{figure*}

\subsection{Mid-Air Ultrasound Tacton Design}
\label{sec:metaphor}

\begin{figure*}[t]
    \centering

    \begin{subfigure}[t]{1.0\textwidth}
    \includegraphics[width=\linewidth]{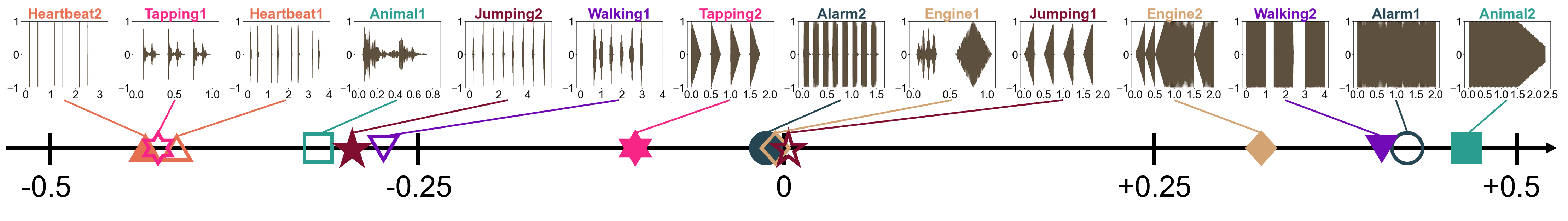}
    \end{subfigure}

    \caption{The first dimension (D1) in the perceptual space of ultrasound Tactons aligned with the rhythmic structure, specifically the number and length of notes or pulses.}
    \label{fig:Trend_UserStudy5_D1}
    \Description{The figure shows signal plots of ultrasound Tactons from Study 5, which are distributed along D1 of the perceptual space. The order of the Tactons from left to right is as follows: Heartbeat2, Tapping1, Heartbeat1, Animal1, Jumping2, Walking1, Tapping2, Alarm2, Engine1, Jumping1, Engine2, Walking2, Alarm1, and Animal2.}
    
\end{figure*}

We created this set based on 14 mechanical Tactons, with two Tactons per seven metaphors: \textbf{Alarm}, \textbf{Animal}, \textbf{Engine}, \textbf{Heartbeat}, \textbf{Jumping}, \textbf{Tapping}, and \textbf{Walking}. Kwon et al.~\cite{kwon2023can} selected these mechanical Tactons from the VibViz vibration library~\cite{seifi2015vibviz} and studied their perceptual dissimilarity.
Since the mechanical Tactons from~\cite{seifi2015vibviz} were not based on a mathematical formula, we could not generate the corresponding mid-air Tactons using the parameter-based method as in Studies 2--4.
Instead, we extracted their temporal envelopes and frequencies from the signals using the Hilbert transform and the short-time Fourier transform (STFT) with a window size of 40\,ms, similar to the window size used in Study 4 (Section \ref{sec:rhythm}).
Then, we applied a 40\,ms moving average filter to the envelopes with pulses shorter than 40\,ms (only \textbf{Heartbeat2} in this set), similar to Study 4.
We additionally used a 10\,ms moving average filter to reduce temporal noise in the extracted envelopes for \textbf{Animal1}, \textbf{Engine1}, \textbf{Jumping2}, and \textbf{Tapping1}.
Finally, we normalized the max amplitude for all the temporal envelopes to 100\%\ amplitude on the ultrasound device to ensure users can perceive the patterns~\cite{howard2019investigating}.
Next, we created the mid-air ultrasound Tactons by multiplying the resulting temporal envelopes with a sinusoid with time-variant AM frequencies and $U(t)$, the ultrasound carrier signal.
This process led to 14 mid-air Tactons corresponding to the mechanical Tactons while ensuring that users can perceive all the pulses in the Tactons (Figure \ref{fig:Signal_UserStudy5}).
These Tactons had complex waveforms (i.e., temporal envelopes and AM frequencies) and their durations ranged from 0.84 to 5.38 seconds. 

\subsection{Participants}
We recruited 15 new participants (two females and 13 males, 18--46 years old), including 2 left-handed, 11 right-handed, and 2 ambidextrous users.
On average, the participants took 71 minutes and received a \$22 USD Amazon gift card.

\subsection{Results}
\textbf{Distinguishability of mid-air ultrasound Tactons:} 
The ultrasound Tactons were primarily structured according to their rhythmic structure (D1), especially the number and length of notes or pulses, and the total duration of the Tacton (D2). 
Along D1, the rhythmic structure varied from Tactons with several short pulses on the left (negative D1) to Tactons with continuous envelopes or long notes on the right (Figure~\ref{fig:PS_UserStudy5}-left and Figure~\ref{fig:Trend_UserStudy5_D1}). 
The position of Tactons along D2 showed a very strong correlation with the total duration of the Tactons ($r = -0.82$, $p<0.001$). 
The three Tactons with higher D2 values, \textbf{Animal1}, \textbf{Engine1}, and \textbf{Tapping1}, had the shortest total durations in the set (Figure~\ref{fig:PS_UserStudy5}-left). 
We could not identify any temporal parameters for the perceptual space in higher dimensions (D3 and D4).
The distances on the D3-D4 perceptual space and the perceptual dissimilarities reported by users showed no relationship, suggesting these dimensions may mainly represent mathematical residuals from the nMDS analysis, rather than having a perceptual basis.

\textbf{Comparison with mechanical vibrations:} 
The perceptual dissimilarities for mid-air ultrasound Tactons showed a strong correlation (Spearman's $\rho = 0.78$, $p<0.001$) with the perceptual dissimilarities for mechanical Tactons from~\cite{kwon2023can}.
The mechanical Tactons were structured according to their rhythmic structure (D1), the change in frequency spectrum (D2), and the total duration of the Tacton (D3).
D1 for mechanical vibrations strongly correlated with D1 on the ultrasound Tactons ($r = 0.97$, $p<0.001$), reflecting the rhythmic structure in both cases.

D2 of mechanical Tactons showed a different tendency from D2 for ultrasound Tactons.
The four Tactons with continuous envelopes or long notes, \textbf{Alarm1}, \textbf{Animal2}, \textbf{Engine2}, and \textbf{Walking2} were also located on the right (positive D1), but their distribution along D2 was farther apart than along D2 of ultrasound Tactons. The amount of change in the frequency spectrum for these Tactons increased when moving from positive to negative D2 values in the perceptual space of mechanical Tactons (Figure~\ref{fig:PS_UserStudy5}-right).
\textbf{Animal2}, with the highest value on D2, had a single frequency component.
In \textbf{Walking2} and \textbf{Engine2}, the frequency gradually increased or decreased over time.
Next, \textbf{Alarm1} alternated between pulses with two distinct frequencies.
Tactons with short pulses (\textbf{Heartbeat1} - \textbf{Heartbeat2} - \textbf{Walking1} - \textbf{Animal1}) had positive values along D2, suggesting users could not fully perceive the frequencies in their short pulses.

Similar to D2 in the perceptual space of ultrasound Tactons, the position of mechanical Tactons along D3 showed a strong correlation with the total duration of the Tactons ($r = -0.61$, $p<0.05$).
The three mechanical Tactons with higher D3 values, \textbf{Animal1}, \textbf{Engine1}, and \textbf{Tapping1}, had the shortest total durations in the set, showing similar tendency in D2 of ultrasound Tactons (Figure~\ref{fig:PS_UserStudy5}). 
We could not identify any temporal parameters aligned with D4 in the perceptual space for mechanical Tactons.

\textbf{Brief discussion:} 
Our results demonstrated that two temporal ultrasound parameters, rhythmic structure and total duration, primarily influenced the perceptual distances of ultrasound Tactons designed using metaphor-based approach. Rhythmic structure and total duration were also present in the perceptual space of mechanical Tactons, providing shared parameters for designing metaphor-based Tactons across contact and contactless interactions. In contrast, frequency variations led to different perceptual distances in ultrasound and mechanical Tactons, practically removing one dimension from the perceptual space of ultrasound Tactons. 
Humans have a more sensitive discrimination threshold for the frequency of mechanical vibrations than for the AM frequency of ultrasound vibrations. 
Thus, ultrasound designers must use larger AM frequency step sizes to create distinguishable ultrasound Tactons.

\begin{table*}[t]
  \begin{tabular}{c>{\centering\arraybackslash}m{0.7cm}*{10}{>{\centering\arraybackslash}m{0.7cm}}}
    \toprule
    STM Combination    & 1 & 2 & 3  & 4  & 5  & \textbf{6}  & 7  & 8  & 9  & 10 \\ \hline
    \midrule
    Radius ($mm$)    & 0 & 5 & 5  & 5  & 10 & \textbf{10} & 10 & 15 & 15 & 15 \\ \hline
    Drawing speed ($m/s$) & 0 & 6 & 12 & 18 & 6  & \textbf{12} & 18 & 6  & 12 & 18 \\ \hline
  \bottomrule
\end{tabular}
\caption{Combinations of STM parameters tested in the preliminary study. Combination 1 with 0\,$mm$ radius and 0\,$m/s$ drawing speed is the pure AM rendering. In studies 2--5, we used the STM parameters in combination 6.}
\label{tab:combinations}
\end{table*}

\begin{figure*}[t]

    \centering
  
    \begin{subfigure}[t]{0.9\textwidth}
    \includegraphics[width=\linewidth]{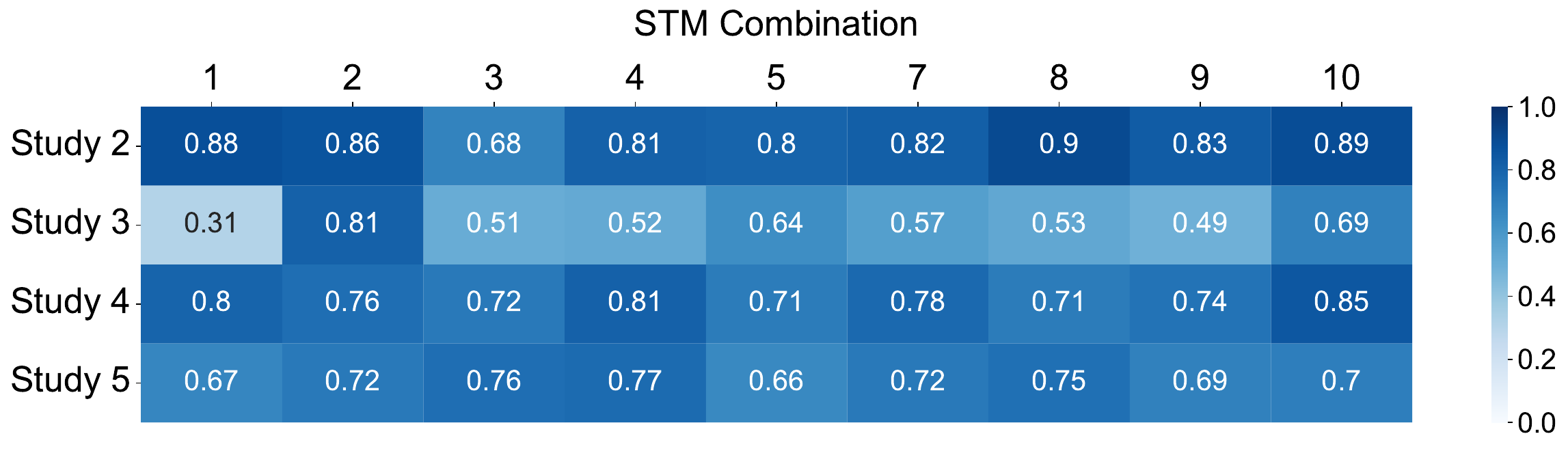}
    \end{subfigure}

    \caption{
    Spearman's rank correlations of dissimilarity values in the nine STM combinations with the reference mid-air ultrasound Tacton sets used in Studies 2--5 (Combination 6). 
    }
    \label{fig:combinations}
    \Description{XX}
    
\end{figure*}

\section{Preliminary Study on Generalizability of Findings} 
\label{sec:preliminary}
In the above studies, we kept a constant radius (10\,$mm$) and drawing speed (12\,$m/s$) for the focal point and focused on identifying the salient temporal parameters for creating distinguishable ultrasound Tactons, but can our results be generalized to other STM radius and drawing speeds?
To answer this question, we tested nine combinations of radius and drawing speed (Table~\ref{tab:combinations}) for the four mid-air ultrasound Tacton sets in Studies 2--5.
Thus, we created a total of 36 Tacton sets, collected two ratings for each Tacton set from six participants, and averaged the ratings to derive dissimilarity spaces for each Tacton set.
The study setup, procedure, and analysis were the same as in Studies 2--5.

For the Tactons in Studies 2, 4, and 5, the dissimilarity ratings and perceptual spaces showed high correspondence across the STM parameters. The Spearman correlations for the dissimilarity values in the nine STM combinations showed strong correlations (mean $\rho$ = 0.77) with the dissimilarity values for the reference Tacton sets in these studies (Figure~\ref{fig:combinations}).
The derived perceptual spaces were also similar with minor differences, capturing the same salient temporal parameters as in the main studies. 
The results suggest that the variations in temporal parameters in Studies 2, 4, and 5 determined the perceived similarities, regardless of the specific values for the two STM parameters.

For the Tacton set in Study 3, the dissimilarity values showed moderate correlations with the dissimilarities in Study 3 (mean $\rho$ = 0.56).
We conjecture that the complex spectrum created by multiple sinusoids and the interaction between AM and drawing frequencies impacted the distinguishability in this Tacton set, leading to lower correlation values than in other Tacton sets.
The derived perceptual spaces were separated by the two temporal parameters of envelope frequency and amplitude, as in Study 3.
However, local distributions formed by the superposition ratio varied across the STM radius and drawing speed.
Moreover, the participants reported that Tactons with the first combination (i.e., pure AM rendering) were too weak to perceive, resulting in low correlation ($\rho$ = 0.31) with Study 3. 
For this pure AM rendering, the envelope frequency divided the perceptual space into two clusters, but amplitude did not impact the perceptual space, perhaps due to the low perceived intensity of the pure AM rendering.
Thus, designers should be careful when designing Tactons with multiple sinusoids and STM parameters, as both factors can create complex spectra and impact similarity perception.

Overall, our results from the main studies correlated with the results from the nine combinations of STM parameters.
These results and the high correlations in similarity ratings between the ultrasound and mechanical Tactons in Studies 2--5 suggest that the temporal factors played a dominant role in Tacton perception. The use of multiple sinusoids in Study 3 resulted in the lowest correlation with mechanical Tactons ($\rho$ = 0.61) among the four Tacton sets which can be due to the impact of STM parameters.
Nevertheless, further studies are needed to fully establish the interaction of AM and STM parameters and their impact on user perception.

\section{Discussion}

In this paper, we designed and conducted five studies.
Study 1 uncovered AM frequency JNDs of 47\%--77\% at 30\,Hz, 80\,Hz, and 210\,Hz, and identified significant differences in JNDs between 30\,Hz and 80\,Hz, and between 30\,Hz and 210\,Hz.
Studies 2--5 explored perceptual spaces for four ultrasound Tacton sets, highlighting salient temporal parameters influencing perceptual dissimilarity: envelope frequencies ($\leq$ 5\,Hz), two amplitude levels, rhythmic structures including the number and length of notes (i.e., pulses) and their evenness, and total durations.
Studies 2--5 also revealed strong correlations in perceptual dissimilarities between mid-air ultrasound and mechanical Tactons, with Spearman's $\rho$ of 0.70, 0.61, 0.89, and 0.78, respectively.
Based on these results, we present guidelines for designing distinguishable ultrasound Tactons and outline implications for future work.

\subsection{Design Guidelines for Distinguishable Mid-Air Ultrasound Tactons}
We compiled six guidelines that can support both parameter-based and metaphor-based design approaches.
Prior work suggest that tactile acuity decreases by about 1\% per year from age 12 to 85~\cite{lederman2009haptic}.
Thus, these guidelines should be carefully applied when the target population is significantly older than our participants (18--46 years old, mean: 24.2).

\textbf{1. Use low envelope frequency ranges ($\leq$ 5\,Hz) to create distinct ultrasound Tactons.}
Study 2 demonstrated that low envelope frequencies can provide a distinct tactile sensation but the perceptual distance dropped drastically when the envelope frequency exceeded 5\,Hz. A similar drop was reported for mechanical Tactons in prior work~\cite{park2011perceptual} but at higher envelope frequencies (40\,Hz).
Also, in Study 3, changes to the envelope frequency (0\,Hz vs. 4\,Hz) showed the largest effect on the perceived similarity of Tactons compared to variations in amplitude, AM frequency, and superposition ratio.
Based on the results, we conclude that an envelope frequency lower than or equal to 5\,Hz enables users to distinguish between Tactons clearly.

\textbf{2. Two amplitude levels can be used but with caution.}
Study 3 showed that users could distinguish between Tactons with two amplitude levels (half and full) on the mid-air ultrasound device, perhaps due to the gap in their sensation levels.
The preliminary study also suggested that amplitude was more salient than AM frequency and superposition ratio across STM parameters, except when using pure AM rendering.
The influence of amplitude on perceptual dissimilarity decreased when used with rhythmic structure in Study 4.
These results suggest that the saliency of amplitude depends on the other temporal parameters and designers must be careful when using multiple temporal parameters to create Tactons.
In addition, designers must select the amplitude with care because the intensity of mid-air ultrasound technology is often low~\cite{rakkolainen2020survey, freeman2019haptiglow}, and the amplitude detection thresholds can vary across the AM frequency spectrum and modulation techniques~\cite{howard2019investigating, hasegawa2018aerial}.

\textbf{3. The number and length of notes (or pulses) and their evenness are key attributes for the perceptual dissimilarity of ultrasound Tactons.}
Among the myriad of temporal parameters in Tactons, the rhythmic structure, particularly the number and length of pulses and their evenness, greatly influenced the perceptual dissimilarity spaces of both parameter-based and metaphor-based mid-air ultrasound Tactons (Studies 4--5).
The correlations in perceptual dissimilarities between ultrasound Tactons that varied in rhythmic structure and their corresponding mechanical Tactons were high (mean Spearman's $\rho$ = 0.84) in Studies 4--5. 
Moreover, the number and total length of notes strongly correlated with the two perceptual dimensions of rhythmic ultrasound Tactons ($r = 0.75$ and $r = -0.94$, respectively).
Therefore, designers can use rhythmic structure in both paramber-based and metaphor-based approaches to create distinguishable Tactons for users across contact and contactless technologies.

\textbf{4. Ensure different total durations for complex Tactons, especially when their rhythmic structures are similar.}
Besides the rhythmic structure, the total Tacton duration also aligned with the second perceptual dimension for ultrasound Tactons ($r = -0.82$) in Study 5. Thus, Tacton duration can be an important design parameter for creating distinguishable Tactons.
When two Tactons had the same duration and number of pulses (e.g., 2 seconds, 4 pulses), they were perceived as very similar, despite having different pulse envelopes (e.g., ramp-up vs. ramp-down pulses).
When the total durations were close, this trend remained consistent even for ultrasound Tactons with slightly varied rhythmic structures (6--8 pulses).

\textbf{5. Ensure large step sizes when selecting AM frequencies.}
Study 1 results showed high JND values for AM frequencies (47\%--77\%).
Also, for complex Tactons in Study 5, users reported high perceptual similarity for Tactons when their time-variant AM frequencies did not significantly exceed JNDs.
In contrast, the differences in time-variant frequencies of mechanical Tactons significantly exceeded JNDs, and the frequency served as a perceptual dimension for complex mechanical Tactons.
These findings suggest that a larger step size than the JND in AM frequency is essential when designing ultrasound Tactons, especially when using a chirp pattern or alternating between different AM frequencies.
For designers aiming to provide a perceptual frequency variation in ultrasound Tactons similar to frequency changes in mechanical Tactons, one possibility is to use a larger AM frequency spectrum by considering the relative JND ratio between ultrasound and mechanical Tactons. 
We hypothesize this approach may provide distinguishable ultrasound Tactons varying in AM frequency spectrum akin to mechanical vibrations. This hypothesis must be tested in future work.

\textbf{6. Superposition ratio can provide qualitatively different sensations, but the impact can depend on the STM parameters.}
Study 3 revealed that the superposition ratio contributed less to perceptual dissimilarity than amplitude and envelope frequency.
In the perceptual space, positions of Tactons superimposed with an equal ratio of low and high AM frequencies (M) consistently fell outside a line connecting the positions of the Tactons with low and high AM frequencies (L and H).
This result suggests that an ultrasound Tacton designed using a superposition ratio may provide qualitatively different sensations from Tactons designed with single AM frequencies.
Thus, the superposition ratio may offer a design opportunity for designers seeking to create complex ultrasound Tactons using multiple sinusoids.
On the other hand, our preliminary study with various STM combinations showed lower correlation in Study 3 (mean Spearman's $\rho$ = 0.56), suggesting that the effect of the superposition ratio can depend on STM parameters.
Therefore, when using a superposition ratio and STM parameters for ultrasound Tactons, designers must consider both the spectral components from the drawing frequency and the superposition of the multiple sinusoids.

\subsection{Implications for Future Work}
This paper contributes new data on the just noticeable differences (JND) and perceptual distinguishability of mid-air ultrasound Tactons. We outline how our results can inform future research and haptic design practices with this technology.

\textbf{Designers can utilize our results to improve the distinguishability of spatiotemporal mid-air ultrasound Tactons for users.}
A key advantage of mid-air ultrasound technology is that it allows haptic designers to create spatial patterns in space. Thus, prior studies have generated a variety of haptic shapes with the technology and assessed their identification accuracy to find shapes that users can discern easily.
In contrast, we investigated the efficacy of temporal parameters on the distinguishability of ultrasound Tactons when the focal point is moving along a fixed path (i.e., a 2D circle). Future work can combine our results with prior work to design 2D/3D haptic shapes that vary on their temporal parameters to make the shapes easier to distinguish for users~\cite{mulot2021dolphin, howard2019investigating, mulot2023improving}.

\textbf{Researchers can integrate our findings into graphical haptic design tools, to facilitate the creation of ultrasound Tactons for haptic designers.}
Given the complexity of designing for ultrasonic mid-air haptics, recent studies have proposed graphical user interfaces to support mid-air ultrasound novices and experts in creating Tactons for various applications and scenarios~\cite{mulot2021dolphin,theivendran2023rechap, seifi2023feellustrator}.
Our studies identified which temporal parameters are effective for creating distinguishable mid-air ultrasound Tactons and how their relative saliency changes when combined with other parameters.
Thus, researchers can incorporate our findings into ultrasound design tools, e.g., in the form of sensation palettes, to facilitate the design of distinguishable Tactons for mid-air ultrasound designers.

\textbf{Our results can inform the development of future computational models to predict similarity perception for mid-air ultrasound Tactons.}
Investigating the perceptual space of Tactons is both time-consuming and expensive.
Conducting a pair-wise rating study requires collecting similarity scores for all possible pairs in a set to derive a perceptual space, leading to $O(n^2)$ comparison tasks. 
To accelerate the process of designing distinguishable mechanical Tactons, Lim and Park proposed a computational model to predict perceptual distances of a given mechanical Tacton set by simulating the neural transmission from the mechanoreceptors in the skin to the brain~\cite{lim2023can}.
The model was trained on similarity data from prior literature.
Our paper provides similarity data for four sets of mid-air ultrasound Tactons, including seven temporal parameters and their combinations, and compares the results with mechanical Tactons.
Thus, our study supplies new data for training similar perceptual models for ultrasound technology or expanding the scope of existing computational models from mechanical vibrations to mid-air ultrasound vibrations.

\textbf{Future work should study the impact of ultrasound's temporal parameters on the perceived intensity by users.} 
First, when studying JNDs for three AM frequencies of mid-air ultrasound, we purposefully did not match the perceived intensities for these frequencies to make our results practically relevant to haptic designers who do not match the perceived intensities when designing Tactons.
Future psychophysics studies can complement our results by examining a wider range of AM frequencies and matching their perceived intensities to provide a perceptual basis for mid-air ultrasound stimulation. 
Second, we observed that the perceptual distance drastically decreased when the envelope frequency increased from 5\,Hz to 10\,Hz.
The lower perceived intensities of ultrasound Tactons may be the reason behind their smaller perceptual distances compared to mechanical Tactons~\cite{hwang2010perceptual}. Future studies can investigate the relationship between envelope frequency and perceived intensity.
Finally, future work can further explore the superposition of multiple AM frequencies for mid-air ultrasound vibrations.
Recent research has investigated the impact of various superposition ratios of mechanical vibrations on perceptual distinguishability~\cite{hwang2017perceptual} and the perceived intensities~\cite{yoo2022perceived}.
Such studies provide important insights for designing complex mechanical Tactons and can inspire similar investigations for mid-air ultrasound vibrations.

\section{Conclusion}

The multitude of design parameters in mid-air ultrasound haptics offer new opportunities and challenges for designers.
Our work aims to support haptic designers by charting the distinguishability of temporal parameters in mid-air ultrasound Tactons and comparing the results to the user perception of mechanical Tactons.
We proposed the design guidelines for the ultrasound Tactons based on lab studies. 
In the future, designers and practitioners can further verify these guidelines once the Tactons are integrated into end-user applications. 
We hope our findings help designers create perceptually salient Tactons for contactless interactions and inspire future research to uncover the underlying perceptual mechanisms for intuitive and unique tactile experiences for end-users.


\newpage
\newpage

\begin{acks}
We would like to thank Seungmoon Choi for lending us the mid-air ultrasound haptic device, and Charan Reddy Guthi and Kevin John for assisting with the pilot tests.
We thank all participants and reviewers for their valuable input on the project. 
This work was supported by research grants from VILLUM FONDEN (VIL50296), the National Science Foundation, the Institute of Information \& communications Technology Planning \& Evaluation (IITP) funded by the Korea government (MSIT) (No.2019-0-01842, Artificial Intelligence Graduate School Program (GIST)), and the Culture, Sports and Tourism R\&D Program through the Korea Creative Content Agency funded by the Ministry of Culture, Sports and Tourism in 2023 (RS-2023-00226263).
\end{acks}

\bibliographystyle{ACM-Reference-Format}
\bibliography{99_Bibliography}











\end{document}